%% file: main.tex
\documentclass{stile/aa}

\input{include_tex/pacchetti}
\usepackage{footmisc}
\input{include_tex/aa_url_patch}

\input{include_tex/journals}

\input{include_tex/definizioni}

\input{include_tex/additional_def}
\input{include_tex/colors}

\graphicspath{{figures/}, {figures/evolution/}, {figures/maps/}, {figures/profiles/}}

\def\cloud{{\rm MC}}
\defcitealias{Sommovigo:2020}{S20}
\defcitealias{Decataldo2020}{D20}
\defcitealias{Jaquez-Dominguez2023}{JD23}

\begin{document}

\title{Investigating the ultraviolet and infrared radiation through the turbulent life of molecular clouds}
\titlerunning{UV and FIR in MCs}

\author{
Fabio Di Mascia \orcidlink{0000-0002-9263-7900} \inst{1}\fnmsep\thanks{\href{mailto:fabio.dimascia@sns.it}{fabio.dimascia@sns.it}} \and
Andrea Pallottini \orcidlink{0000-0002-7129-5761} \inst{1} \and
Laura Sommovigo \orcidlink{0000-0002-2906-2200} \inst{2} \and
Davide Decataldo \orcidlink{0000-0002-7280-8178} \inst{3}
     }
\authorrunning{Di Mascia et al.}
\institute{
Scuola Normale Superiore, Piazza dei Cavalieri 7, 56126 Pisa, Italy \and
Center for Computational Astrophysics, Flatiron Institute, 162 5th Avenue, New York, NY 10010, USA
\and
Institute of Theoretical Astrophysics, University of Oslo, PO Box 1029, Blindern 0315, Oslo, Norway
}
\date{Received 25 June, 2024; accepted XX XX, XXXX}

\abstract
%
%
{
Molecular Clouds (MCs) are the place where stars are formed and their feedback starts to take place, regulating the evolution of galaxies. Therefore, MCs represent the critical scale at which to study how ultra-violet (UV) photons emitted by young stars are reprocessed in the far-infrared (FIR) by interaction with dust grains, thereby determining the multi-wavelength continuum emission of galaxies.
}
%
%
{
Our goal is to analyze the UV and IR emission of a MC at different stages of its evolution and relate its absorption and emission properties with its morphology and star formation rate. Such a study is fundamental to determine how the properties of MCs shape the emission from entire galaxies.
}
%
%
{
We consider a radiation-hydrodynamic simulation of a MC with self-consistent chemistry treatment. The MC has a mass $M_{\rm MC} = 10^5~\msun$, is resolved down to a scale of $0.06\, \rm pc$, and evolves for $\simeq 2.4$~Myr after the onset of star formation.
We post-process the simulation via Monte Carlo radiative transfer calculations to compute the detailed UV-to-FIR emission of the MC. Such results are compared with data from physically-motivated analytical models, other simulations, and observations.
}
%
%
{%
We find that the simulated MC is globally UV-optically thick, but optically-thin channels allow for photon escape ($0.1\%-10\%$), feature which is not well-captured in the analytical models. 
Dust temperature spans a wide range ($T_{\rm dust} \sim 20-300$~K) depending on the dust-to-stellar geometry, which is reproduced reasonably well by analytical models.
However, the complexity of the dust temperature distribution is not captured in the analytical models, as evidenced by the 10 K (20 K) difference in the mass (luminosity) average temperature. Indeed, the total IR luminosity is the same in all the models, but the IR emission peaks at shorter wavelengths in the analytical ones.
Compared to a sample of Galactic clouds and other simulations, our Spectral Energy Distribution (SED) is consistent with mid-IR data, but peaks at shorter wavelengths in the IR. This is due to a lack of cold dust, as a consequence of the high gas -- and thus dust -- consumption in our simulated MC.
The attenuation properties of our MC change significantly with time, evolving from a Milky-Way-like to a flatter, featureless relation.
On the IRX-$\beta$ plane, the MC position strongly depends on the line of sight and on its evolutionary stage. When the MC starts to disperse, the cloud settles at $\log(\rm IRX) \sim 1$ and $\beta \sim -0.5$, slightly below most of the local empirical relations.
}
%
%
{
This work represents an important test for MC simulations and a first step toward the implementation of a physically-informed, sub-grid model in large-scale numerical simulations to describe the emission from unresolved MC scales and its impact on the global galaxy SED.
} 


\maketitle

\section{Introduction}

Molecular Clouds (MCs) are the birthplace of stars. Gas in overdense regions within MCs cools and collapses until the density is high enough to trigger star formation (SF). This process is regulated by the interplay between pressure support from turbulence motions and the gravitational force, which enables the formation of filaments and clumps \citep[e.g.][]{Brunt2009, Krumholz2012ApJ...745...69K, Colombo2014a, Dobbs2015MNRAS.447}. 
The newly formed stars emit photons which heat and ionize the surrounding interstellar medium (ISM), altering the molecular gas reservoir available for star formation \citep[e.g.][]{Whitworth1979}. Further, $M_\star > 10~\msun$ stars affect the ISM via stellar winds \citep[e.g.][]{Castor1975ApJ, Weaver1977ApJ} and then they end their lives as supernovae (SNe) explosions after $\sim 10$~Myr \citep[e.g.][]{Sedov1958, Ostriker1988}, injecting thermal energy and momentum into the ISM. 
The effect of stellar feedback is essential in determining the complex structure of MCs, their evolution, and ultimately the star formation efficiency (SFE or $\epsilon_\star$), i.e. the fraction of gas converted into stars during the MC lifetime\footnote{Throughout this paper we define the star formation efficiency as the total initial gas mass converted into stars at the end of the cloud lifetime, e.g. $\epsilon_\star = M_\star(t=t_{\rm end})/M_\cloud$.}.

Observations of Molecular Clouds (MCs) in the Milky-Way suggest low values of the star formation rate (SFR) and consequently SFE of only a few percent \citep{Evans2009ApJS..181..321E, Forbrich2009ApJ...704..292F, Garcia2014ApJS..212....2G, Louvet2014, Chevance2020MNRAS.493.2872C}, with the most active clouds being able to convert $\sim 10 \%$ of their gas mass into stars \citep[e.g.][]{Lee2016ApJ...833..229L}. 
These observations probe the dust grains in the cold, dense gas within the MCs, which obscure optical starlight and re-emit it at longer wavelengths, from the near-infrared (NIR) to the sub-mm band. Through this emission, infrared (IR) surveys with Spitzer and Herschel \citep[e.g.][]{Churchwell2009PASP, Molinari2010} have provided the first view of the clumps and filaments in the Milky Way plane, providing a picture of SF in our Galaxy.

From a theoretical point of view, MCs can be studied by analytical means \citep[e.g.][]{Whitworth1979, Williams1997ApJ...476..166W, Matzner2002ApJ, Krumholz2012ApJ...745...69K, Vazquez-Semadeni2019} which are able to pinpoint the relevant physical process responsible for the MC evolution, and numerical simulations, which can account for the complex structure of MCs and better constrain the interplay of star formation and stellar feedback \citep[e.g.][]{kim:2017,jeffreson:2024}, for instance focusing on the impact of radiation \citep[e.g.][]{Decataldo2020,menon:2023}, stellar winds \citep[e.g.][]{geen:2021,lancaster:2021,lancaster:2021_b}, the role of magnetic field in shaping the MC structure \citep[e.g.][]{federrath:2012,Hennebelle2022}, or the influence of the galactic (large scale) environment on the cloud formation \citep[e.g.][]{butler:2015ApJ,Smith2020MNRAS.492.1594S}.

A fundamental ingredient in numerical simulations is the modelling of radiation from stars. In particular, $M_\star > 5~\msun$ stars emit a copious amount of photons in the far-ultraviolet (FUV; $6~{\rm eV} < h\nu < 13.6~{\rm eV}$), which can dissociate CO and H$_2$ molecules, and in the extreme ultraviolet (EUV; $h\nu > 13.6~{\rm eV}$), which ionize hydrogen and helium.
The difference in pressure between the hot, ionized gas, and the cold, molecular ISM, leads to the propagation of shocks ahead of dissociation/ionization fronts, compressing the gas and driving turbulent motions \citep[e.g.][]{Kahn1954, Williams2018}, possibly opening channels through which Lyman continuum and Lyman$\alpha$ photons can leak out of the MC \citep[e.g.][] {Kakiichi2021ApJ...908...30K}.
While dense clumps lose their mass because of gas photo-evaporation, at the same time, the radiation-driven compression can also trigger star formation \citep[e.g.][]{Kessel-Deynet2003, Dale2005MNRAS.358..291D, Bisbas2011, decataldo:2019, Decataldo2020}. An accurate modeling of this complex interplay between stellar radiation and gas dynamics is therefore essential to investigate the SFR within a MC.

However, despite tremendous recent progresses, there are still limitations in current theoretical models. Radiation hydrodynamic (RHD) simulations implement RT on-the-fly with only a handful of energy bins, because of the high computational costs \citep[e.g.][]{Decataldo2020}. Therefore, only a small portion of the total spectral energy distribution (SED) of the MCs is accessible, and with a poor wavelength resolution.
Moreover, the dust treatment is rarely included in these simulations, so radiation reprocessing by dust grains is usually not accounted for. As MCs are typically observed in IR bands \citep[e.g.][]{Lin2017ApJ...840...22L, BinderPovich2018ApJ...864..136B, Ladjelate2020, Potdar2022MNRAS.510..658P}, it is essential to have a tool to model the full UV-to-FIR SED, including the dust contribution, in order to effectively benchmark theoretical models against observations.

A common framework to achieve this task is to compute the propagation of radiation in post-process in order to produce synthetic observations to be directly compared with observations, a strategy widely adopted in the context of galaxy environment simulations \citep[e.g.][]{Camps2016, Koepferl2017ApJ...849....3K, Behrens:2018, Haworth2018NewAR..82....1H, Cochrane2019MNRAS, Reissl2020A&A...642A.201R, DiMascia2021infrared}. 
Such synthetic observations can serve multiple purposes. They can be used to test the robustness of hydrodynamic simulations to reproduce the structure and physical properties of MCs and, conversely, they can also be used to quantify the limits of current observational techniques in inferring the physical properties of MCs. 

Recently, \citet[][hereafter \citetalias{Jaquez-Dominguez2023}]{Jaquez-Dominguez2023} performed RT calculations in post-process and produced synthetic observations of the MCs formation simulations presented in \citet{Zamora2019MNRAS.487.2200Z}, featuring a young star cluster with massive star formation and ionization feedback.
They explored the evolution of the SED as the cloud morphology changes due to the effect of ionizing photons, which lead to expansion of HII regions and formation of cavities.
These results emphasize how complex, varied and sensitive to the line of sight is the galaxy emission at scales below $\approx 10$~pc. These spatial scales are currently unresolved in cosmological hydro-dynamic simulations of galaxy formation, both large-scale \citep{Schaller2015, Feng2016MNRAS, Pillepich2018, Kannan2022} and zoom-in \citep{Ceverino_Glover_Klessen_2017, Hopkins2018MNRAS.480..800H, Lovell2021MNRAS, pallottini:2022}. Therefore, most of the observable predictions derived from these simulations (e.g. UV escape fraction, UV luminosity function, IRX-$\beta$ relations, etc.) are limited by the lack of resulution on small scales.

In this work, we aim at taking a first step toward improving the description of UV and IR continuum emission in simulations of the ISM at the scale of molecular clouds. We do so by comparing the results of post-processed RHD simulations of a local MC with physically-informed analytical models, with the long-term goal to implement them as sub-grid models in numerical simulations. 
We investigate the multi-wavelength emission of local MCs making use of the RHD simulation performed in \citet[][hereafter \citetalias{Decataldo2020}]{Decataldo2020}. This simulation follows the evolution of a MC, implementing on-the-fly RT with 10 radiation bins and a chemical network including H$_2$ formation and destruction, in order to self-consistently model star formation and stellar feedback. 
We perform RT calculations in post-process using the Monte Carlo RT code \code{SKIRT} \citep{CampsBaes2015, CampsBaes2020} in order to study the multi-wavelength emission from the MC, the UV escape fraction, the dust properties (e.g. dust attenuation and dust grain temperature distribution) and how the MC physical properties (e.g. star formation, morphology) impact its emission throughout its evolution. 
We also make use of the analytical models of MC emission by \citet[][hereafter \citetalias{Sommovigo:2020}]{Sommovigo:2020}, in order to explore how simple physically-motivated prescriptions capture the MC properties of modern RHD simulations. This is a starting point toward the development of a model to describe the structure and the emission of unresolved structures on sub-grid scales ($<1$~pc) in galaxy simulations. 

The paper is organized as follows. In Section \ref{sec:model} we describe the numerical and theoretical models adopted. In particular, in Section \ref{sec:MC_model} we present the hydrodynamic simulations; in Section \ref{sec:RT_model} we illustrate the details of the post-process RT setup; in Section \ref{sec:S20_model} we describe the analytical model from \citetalias{Sommovigo:2020}. In Section \ref{sec:ref_cloud} we analyze in details the properties of the MC at a specific snapshot, which we consider as a reference MC. In Section \ref{sec:evolution} we study the evolution of the physical and emission properties of the MC, and finally we conclude in Section \ref{sec:conclusions}.

\section{Model}\label{sec:model}

\subsection{Molecular Cloud evolution}\label{sec:MC_model}

The radiation-hydrodynamic MC model is fully described in \citetalias{Decataldo2020} and summarized here.

\subsubsection*{Simulation setup}

Simulations are carried out using a customised version of the adaptive mesh refinement code \code{ramses} \citep{Teyssier2002}, which uses a second-order Godunov scheme for the gas hydro-dynamics and a particle-mesh solver for particles as stars, solving self-gravity with a multi-grid solver \citep{guillet:2011}. In \citetalias{Decataldo2020}, the full box sixe is 60 pc, and the resolution range from $\Delta x\simeq 0.9$ pc to $\Delta x\simeq 0.06$ pc, according to a semi-Lagrangian strategy.

Radiation is evolved with the \code{ramses-rt} module \citep{Rosdahl2013MNRAS} by adopting a first-order Godunov solver with a M1 closure relation \citep{aubert:2008}. In \citetalias{Decataldo2020}, 10 radiation energy bins are tracked, covering the transition energy of the 9 photo-reactions included in the chemical network. Note that \citetalias{Decataldo2020} uses a reduced speed of light approximation, $c_{\rm red}=10^{-3}\,c$, which yields an inaccurate propagation of the ionization front when its speed is larger then $c_{\rm red}$, that in the simulation happens only close to very massive stars \citep[cfr.][]{deparis:2019,decataldo:2019,pallottini:2022}.

Non-equilibrium chemistry is accounted for by using \code{krome} \citep{grassi:2014}, with a chemical network accounting for $9$ species (H, $\mathrm{H}^{+}$, $\mathrm{H}^{-}$, $\mathrm{H}_2$, $\mathrm{H}_2^{+}$, He, $\mathrm{He}^{+}$, $\mathrm{He}^{++}$ and free electrons), $46$ reactions (see \citealt{bovino:2016,pallottini:2017} for details), and fully coupled with the \code{ramses-rt} module \citep{pallottini:2019,decataldo:2019}

Initially, the cloud has a mass $M_\cloud = 10^5\,\msun$, a radius $R_\cloud=20$ pc (average number density of $n=120\,\cc$), a constant $Z = \zsun = 0.02$ metallicity and a dust-to-metal ratio of $f_{d}=0.3$ ($M_{\rm dust} = 6.4 \times 10^2~\msun$, see in Sec. \ref{sec:RT_model} for dust properties), i.e. appropriate for a MC in the MW \citep[e.g.][]{heyer:2009}. The MC has a turbulent velocity field set up such that the virial parameter $\alpha = 5 \,v_{\rm rms}^2 \, R_\cloud / G\,M_\cloud=2$, i.e. it is marginally unstable. Further, the MC is embedded in a low density medium ($n=1\,\cc$) set in pressure equilibrium. These initial conditions are evolved for 3 Myr before the cloud is allowed to form stars.

\subsubsection*{Star formation and stellar feedback}

\begin{figure}
    \centering
    \includegraphics[width=0.49\textwidth]{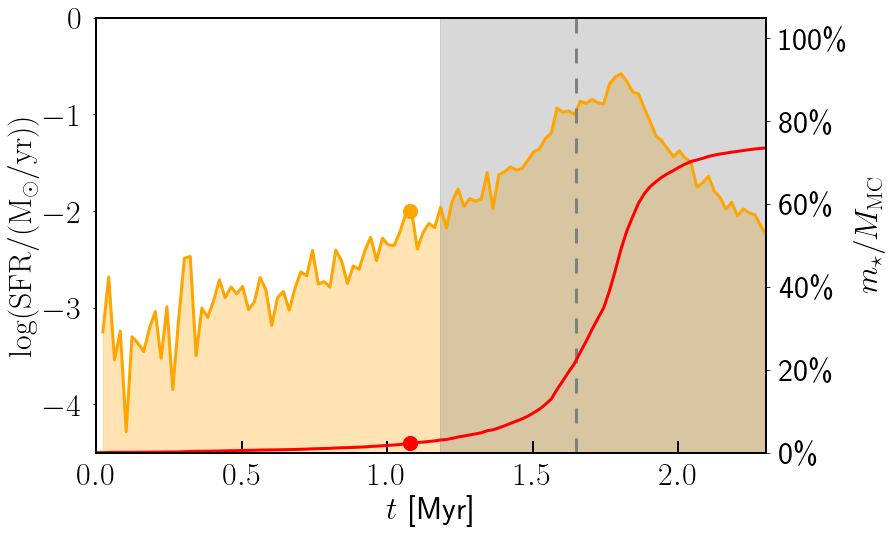}
    \caption{
    Star formation rate (SFR, \emph{orange}) and instantaneous star formation efficiency $\epsilon_\star(t) = M_\star(t)/M_\cloud$, \emph{red}) as a function of time ($t$) for the Molecular Cloud (MC) from \citetalias{Decataldo2020}. At the end of the simulation, the total mass in stars is $M_\star = 7.4 \times 10^4~\msun$.
    The circles mark the time corresponding to the \textit{reference} MC (as define in Sec.~\ref{sec:ref_cloud}), at $t\simeq 1.1$~Myr.
    The dashed vertical line indicates the moment when the MC in \citetalias{Decataldo2020} starts to photo-evaporate.
    The shaded area highlights the period when the cloud is dispersed according to the radiation-pressure model from \citetalias{Sommovigo:2020} (cfr. with Sec.~\ref{sec:S20_model}.) 
    \label{fig:sfr_history}
    }
\end{figure}

In \citetalias{Decataldo2020}, stars can form according to a local \citet{Schmidt1959ApJ,Kennicutt1998ApJ} relation; for each star, the mass ($m_\star$) is drawn from a pre-computed \citet{Kroupa2001MNRAS} initial mass function, where stars with $m\leq 1\,\msun$ are removed for computational efficiency. In \citetalias{Decataldo2020}, the star formation rate (SFR) and the gas mass within the MC are corrected \emph{a posteriori} to take into account the low-mass stars excluded in the simulation.
While including small mass stars can slightly alter gas dynamic, their feedback is expected to be sub-dominant, particularly regarding their contribution to the UV radiation, and therefore to the dust heating (see Sec. \ref{sec:RT_model} and in particular footnote \ref{footnote:uv_estimate_low_mass_stars}).
For this reason throughout this paper we adopt the non-corrected SFR and we consider only the stars actually spawn in the hydrodynamic simulation as radiation sources in the radiative transfer (RT) post-process. 

In the simulation, stars act as radiation point sources and, if massive enough, can inject energy and mass via winds. \citetalias{Decataldo2020} adopts evolutionary tracks from \code{parsec} \citep{bressan:2012} and atmospheric models from \citet{castelli:2003}, in order to compute mass loss rate, wind kinetic power, bolometric luminosities, and spectral shapes. The wind feedback scheme allow for energy and momentum injection in the cells surrounding massive stellar particles \citep{gatto:2017,haid:2018}.

In Fig.~\ref{fig:sfr_history} we show the SFR in the \citetalias{Decataldo2020} MC as a function of time. Note that we do not account for the low-mass stars correction to the SFR in the plot (see Sec.~2.2 \citetalias{Decataldo2020} for details). The SFR increases from $\simeq 10^{-3}~\msunyr$ to a peak of $0.15~\msunyr$ at $t\simeq 1.8\,\rm Myr$; at $t\simeq 1.7$~Myr the cloud starts to photo-evaporate (vertical dashed line in Fig.~\ref{fig:sfr_history}), i.e. the gas start to be ejected from the cloud by winds and radiation pressure, quenching the SFR few Myr afterward. At this stage, it has a stellar mass of $M_\star = 2.2\times 10^4~\msun$ ($M_\star = 4.3 \times 10^4~\msun$ if accounting also for low-mass stars).
Note that the SFR is particularly bursty at early times, with stochastic variation up to 1 dex, and the amplitude of the fluctuations becomes negligible around the $t\simeq 1.5\,\rm Myr$ mark \citep[cfr.][for SFR variations in galaxies]{furlanetto:2022,pallottini:2023, Sun2023ApJ...955L..35S}.

\subsection{Radiative transfer through dust in MCs}\label{sec:RT_model}

The on-the-fly RT module included in the simulations by \citetalias{Decataldo2020} allows to model self-consistently the propagation of radiation and the gas chemical evolution. However, the number of energy bins is too low to make solid predictions for the SED of the MC.
For this reason, we post-process the MC with 3D-radiative transfer calculations by adopting the version 9 of the Monte Carlo code \code{SKIRT}\footnote{\url{https://www.skirt.ugent.be}} \citep{CampsBaes2015, CampsBaes2020}. 

\code{SKIRT} includes a full treatment of absorption and scattering of light by dust grains, and it calculates self-consistently the dust thermal emission. The code can handle arbitrary three-dimensional geometries for both the radiation sources and the dust component, and it includes a large variety of emission models for the sources. Further, \code{skirt} has been widely used to generate precise synthetic observations of simulated galaxies \citep[e.g.][]{Camps2016, Trayford2017MNRAS, Behrens:2018, Cochrane2019MNRAS, Liang:IRXbeta, Vijayan2022FLARESIII}, AGN-hosts \citep[e.g.][]{viaene:2020,DiMascia2021infrared,dimascia:2023}, and recently it has also been applied to simulations of Galactic molecular clouds (see \citetalias{Jaquez-Dominguez2023}).

A RT calculation with \code{SKIRT} requires two main ingredients: a source component, which determines the intrinsic radiation field within the computational domain, and a dust component, which alters the radiation via scattering and absorption, and then thermally re-emits photons, reshaping the radiation field. 

\subsubsection*{Stellar emission}

We import the individual stars in the hydrodynamic simulation. Consistently with \citetalias{Decataldo2020}, their SED is computed based on the stellar atmosphere models by \citet{castelli:2003}, which depend on the stellar metallicity $Z_\star$, the surface temperature $T_\star$ and the surface gravity $g_\star$\footnote{The luminosities provided by \citet{castelli:2003} are normalized to a sphere with unit radius, so they have to be multiplied by the surface of the star, i.e. $4\pi R_\star^2$. Therefore the SED of each star is determined by four parameters.}.
We use the stellar evolutionary tracks computed with the \code{PARSEC} code \citep{bressan:2012} to relate the temperature $T_\star$ and radius $R_\star$ of each star to its mass $m_\star$, assuming that all the stars have the same metallicity as the gas, i.e. $Z_\star = Z_{\rm gas} = \zsun$. 
We underline that, by importing only the stars actually included in the hydrodynamic simulation, only radiation from stars with $m_\star>1~\msun$ is considered in the RT calculation\footnote{We expect the contribution of the low-mass stars not included in the RT post-process to be negligible. By assuming a \citet{Kroupa2002} IMF and a scaling $L_\star\propto m_\star^3$ appropriate to approximate the main sequence, the total luminosity of stars with $m_*>1~\msun$ is about three orders of magnitude higher than the low-mass stars.\label{footnote:uv_estimate_low_mass_stars}}. These stars are the ones contributing to the SFR and total stellar mass budget $M_\star$ shown in Fig.~\ref{fig:sfr_history}. 

\subsubsection*{Dust distribution and properties}

We import in \code{SKIRT} the octree structure of the hydrodynamic simulation. We assign to each cell a dust mass content by re-scaling the gas metal mass with a dust-to-metal ratio $f_{\rm d}=0.3$, consistently with the chemistry model adopted in \citetalias{Decataldo2020}. 
In cells with high gas temperature \citep[$T\gsim 10^6\rm K$, see e.g. ][]{DiMascia2021infrared}, grains are efficiently destroyed via thermal sputtering, as collisions with ions are expected to be more frequent. However, we find this effect to be relevant for $\lesssim 0.1\%$ of the total dust mass in our simulations, so we do not apply any dust destruction recipe in pre-process.

For the dust optical properties, we use the Milky-Way (MW) model from \citet{Weingartner:2001} with the visual extinction-to-reddening ratio equals to $R_{\rm V}=3.1$.
The dust mix consists of graphite, silicate and polycyclic aromatic hydrocarbons (PAHs) grains, whose size distribution spans from $3.5 \times 10^{-4}~\mum$ to $10~\mum$. Each component is discretized in $10$ bins for the computation. 

\subsubsection*{SED modelling}

We restrict the simulated spectra to the wavelength range $0.09-1000~\mum$, with a logarithmic spacing. The limits were chosen to include the UV and optical photons for which the dust absorption/scattering cross sections are higher, and all the relevant wavelengths for the primary emission from stars. We did not include photons below $0.09~\mum$ because we expect hydrogen absorption to be more important than dust in that regime. We collect the radiation adopting the same wavelength grid, with $200$ bins.  We use $5 \times 10^6$ photon packets for our simulations, $50\%$ being equally distributed among different sources, $50\%$ being preferentially assigned to higher luminosity sources.

\code{SKIRT} perform the RT computation in two stages. First, it propagates radiation from the primary sources (i.e. stars). The absorbed radiation and the resulting radiation field is then used to compute the temperature of each grain population in each cell in the computational domain\footnote{We also include stochastic heating by individual photons on small dust grains.}.
In the second stage, the dust thermal emission is propagated. This secondary emission can further interact with the dust grains if the density is high enough that the dust optical depth is non-negligible at longer wavelengths. This computation is repeated until a convergence criterion is satisfied. We stop the iteration when the total absorbed dust luminosity changes less than $3\%$ with respect to the previous step.

\subsubsection*{Sample selection}

We select a total of $24$ snapshots to be post-processed, from $t=0.08$~Myr to $t=2.34$~Myr since the beginning of star formation in the cloud, in order to probe all the different stages of the cloud evolution.

We configure in \code{SKIRT} an instrument, which acts as synthetic detector, to collect radiation from the computational domain. We place it at an arbitrary\footnote{The selected distance does not affect the results as long as the angular size subtended to the cloud is small.} distance of $1$~Mpc.
We setup the same wavelength grid adopted in the \code{SKIRT} RT computation. The detector has a field of view of $60 \ {\rm pc} \ \times 60 \ {\rm pc}$, which covers the whole computational box, and $1024 \ \times 1024$ pixels, in order to achieve the maximum physical resolution of the hydrodynamic simulation. 

We place a total of $6$ instruments at different angular positions around the simulated box in order to explore variations in the emission of the simulated cloud with the line of sight. In particular, we choose lines of sight perpendicular to each face of the simulated cubic box. We pick one of them as the reference line of sight to study the cloud emission.

\subsection{An essential model for dust emission in MCs} \label{sec:S20_model}

In this work, we adopt the model from \citetalias{Sommovigo:2020} as the main tool for comparisons. The physically-motivated model for computing dust emission from MCs from \citetalias{Sommovigo:2020} is summarized below, along with the modification needed to have a closer match with the \citetalias{Decataldo2020} MC setup.

In \citetalias{Sommovigo:2020}, MCs are modelled assuming spherically symmetry and are characterized by a mass $M_\cloud$ equal to the Jeans mass with a number density of $n \simeq 122~\cc$, a radius $R_\cloud$ given by the Jeans length, and a supersonic velocity dispersion $\sigma_\cloud$. In the present, $R_\cloud$ and $M_\cloud$ are the same as \citetalias{Sommovigo:2020}, which are close to the values of the MC in \citetalias{Decataldo2020}; $\sigma_\cloud$ is set such that $\alpha=5/3$ and the dust mass matches the one in \citetalias{Decataldo2020}.

Stars are located at the center of the MC and their radiation is absorbed by the dust as it travels outward; to this end, the cloud is divided in optically thin shells, containing a fraction of dust with a \citet{Weingartner:2001} grain size distribution.
As the UV radiation from the stars is absorbed by dust, it emits as a modified black body, with a temperature dependent on the the absorbed luminosity and grain size, assuming local thermal equilibrium (LTE).
While in the original \citetalias{Sommovigo:2020} stars are allowed to form with a \citet{Schmidt1959ApJ}-\citet{Kennicutt1998ApJ} relation and the UV radiation was computed with \code{starburst99} \citep{Leitherer1999ApJS}, we match the UV SED and intensity field with \citetalias{Decataldo2020}.

The system can be solved once the density profile is specified. \citetalias{Sommovigo:2020} adopts either a homogeneous density (\textit{homogeneous} model) or the profile resulting from the impact of radiation pressure due to the presence of $\HII$ regions \citep[][\textit{radiation-pressure} model]{draine:2011}.

\begin{figure*}
    \centering
    \includegraphics[width=0.9\textwidth]{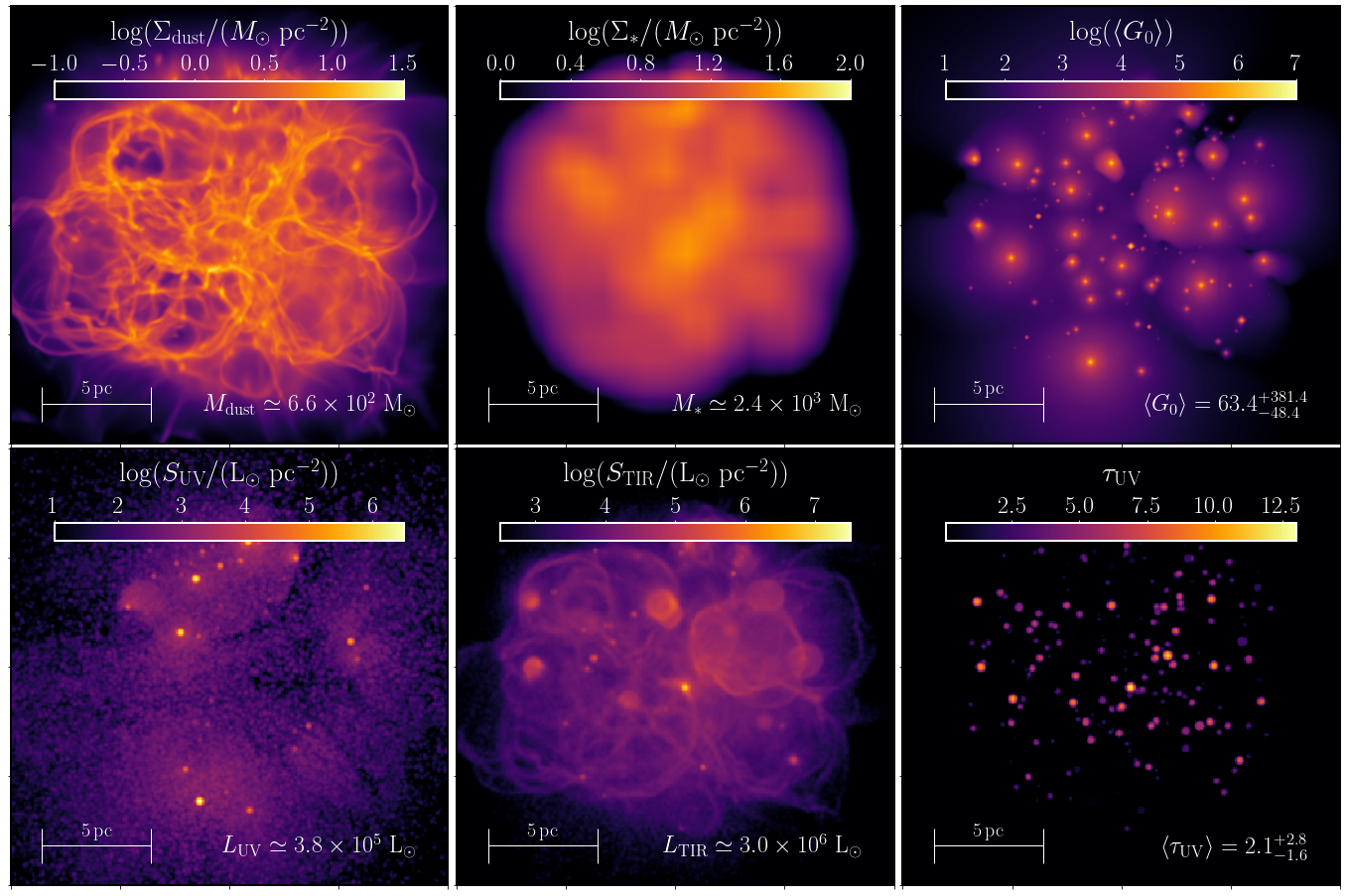}
    \caption{
    Morphology of the reference MC, zooming on the central region of the box with a side of $20$~pc. The maps refer to the reference l.o.s.. 
    Top panels show the dust surface density (left panel), stellar mass surface density (middle panel) and the photon density-weighted Habing flux $G_0$ (right panel). 
    Bottom panels show the UV emission (left panel), IR emission (middle panel) and V-band optical depth (right panel), computed from the post-process RT. 
    The RT maps are smoothed with a Gaussian kernel of $\sigma=1$ pixel with respect to the original resolution.
    For the same diagnostics at different evolutionary times, see Fig.s \ref{fig:cloud_morphology_55}, \ref{fig:cloud_morphology_100}, and \ref{fig:cloud_morphology_135} .
    \label{fig:cloud_morphology}
    }
\end{figure*}

\section{A closer look to prototypical MC} \label{sec:ref_cloud}

As a first step in our investigation of the emission properties of the evolving MC, we focus on a snapshot at $t \approx 1.1$~Myr (after the beginning of star formation), which we will refer to as our \emph{reference} cloud. At this stage, the MC has a (non-corrected) stellar mass of $M_\star \sim 2.3 \times 10^3~\msun$ and a $\SFR \sim 0.01~\msunyr$ (cfr with Fig.~\ref{fig:sfr_history}). 
The choice of this snapshot was motivated by the need to investigate the cloud after star formation has proceeded for enough time for stellar feedback to be effective. Another reason to use this snapshot was that photo-evaporation is over-estimated when using the radiation-pressure model by \citetalias{Sommovigo:2020} with the \citetalias{Decataldo2020} SED as all the stars are placed in the center of the cloud, and the cloud becomes effectively dispersed after $t=1.2$~Myr, making a comparison between the analytic models and the simulated MC less meaningful. Throughout the paper, the time where the cloud is dispersed in the radiation-pressure model is emphasized with a shaded grey area.

We now discuss the cloud morphology at the reference snapshot in Sec.~\ref{sec:cloud_morph}, then its emission properties in Sec.~\ref{sec:cloud_em}, and the dust temperature distribution in Sec.~\ref{sec:dust_temp}.

\subsection{Cloud morphology and dust mass distribution} \label{sec:cloud_morph}

\begin{figure}
    \centering
    \includegraphics[width=0.49\textwidth]{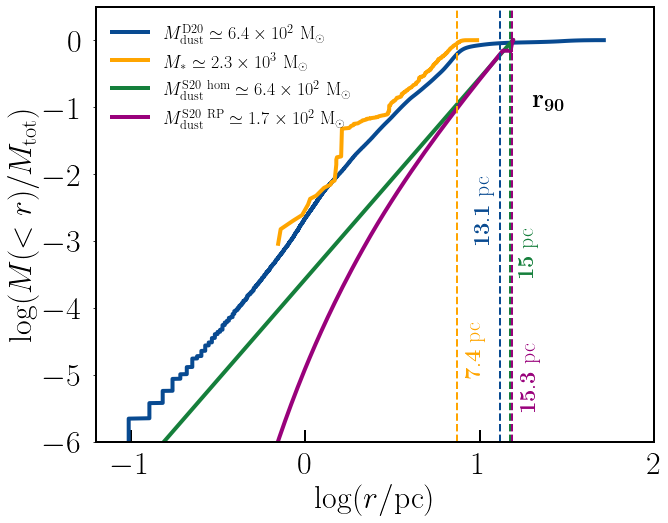}
    \hfill
    \caption{
    Physical properties of the reference MC.
    Fraction of dust mass ($M_{\rm dust}$, blue line) and stellar mass ($M_\star$, orange line) enclosed within a sphere of radius $r$.
    Total masses are indicated in the legend, vertical dashed lines mark the radii containing $90\%$ of the total mass for each component ($r_{\rm dust, 90}$ and $r_{\rm *, 90}$) with values reported in figure.
    As a reference, we show the enclosed dust for the \citetalias{Sommovigo:2020} model in the homogeneous ($M_{\rm dust}^{\rm S20 \ hom}$, green line) and radiation-pressure ($M_{\rm dust}^{\rm S20 \ RP}$, magenta line) case, as well as the radii enclosing the $90\%$ of the total mass in both cases ($r_{\rm dust, 90}^{\rm S20 \ hom}$ and $r_{\rm dust, 90}^{\rm S20 \ RP}$ respectively).
    \label{fig:dust_mass_density_profile}
    }
\end{figure}

\begin{figure}
    \centering
    \includegraphics[width=0.49\textwidth]{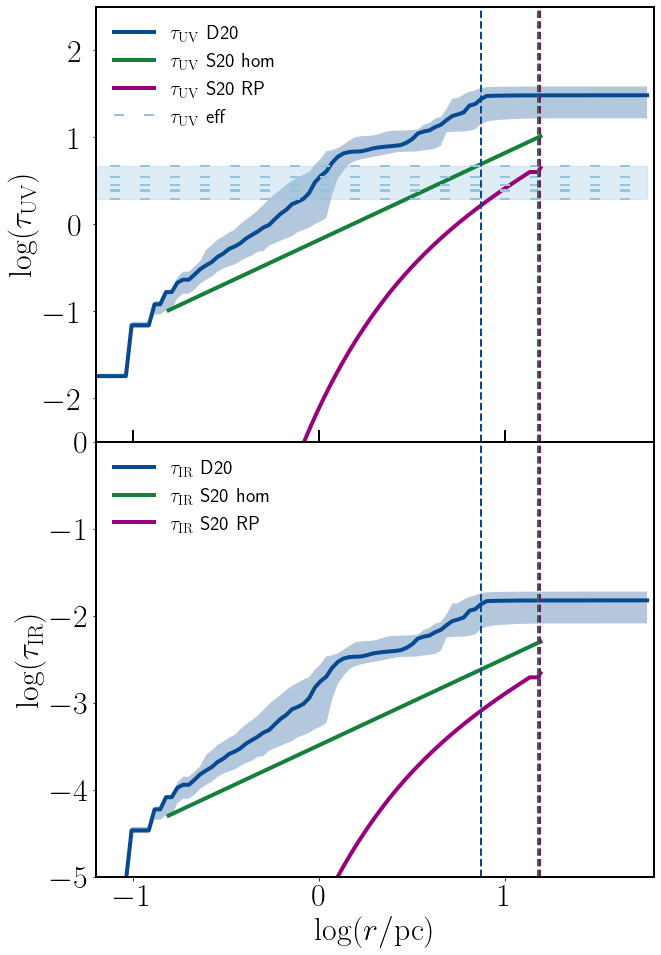}
    \hfill
    \caption{
    Absorption properties of the reference MC.
    \textbf{Top panel.} UV optical depth ($\tau_{UV}$, at $1600~\angstrom~$) as a function of the cloud radius ($r$).
    $\tau_{UV}$ is integrated starting from the center of the cloud. The solid \emph{blue} line gives the median of $\tau_{UV}$ when considering $100$ lines of sight, while the shaded region indicate the variance ($16$th-$84$th percentiles). 
    The shaded \emph{light blue} area brackets the values of the effective optical depth for the six camera instruments, obtained as $F_{\lambda} = F_{\lambda, 0} e^{-\tau_{\lambda, \rm eff}}$, where $F_{\lambda}$ is the detected flux and $F_{\lambda, 0}$ the flux that would be received without dust. We also plot with a dashed line the values of $\tau_{\rm eff}$ for each line of sight.
    The dashed vertical blue line shows the radius at which $\tau_{UV}$ reaches $90\%$ of the total value.
    As a reference, we plot the corresponding optical depth for the \citetalias{Sommovigo:2020} model in the homogeneous case (\emph{green} line) and in the radiation-pressure case (\emph{magenta} line). 
    \textbf{Bottom panel.} Infrared dust optical depth ($\tau_{IR}$, at $100~\angstrom~$) for our reference MC. Similarly to the upper panel we indicated with a \emph{blue} line (shaded region) the median (variance) of the reference cloud.
    The dashed vertical blue line shows the radius at which $\tau_{IR}$ reaches $90\%$ of the total value.
    We also show the results for the \citetalias{Sommovigo:2020} model in the homogeneous case (\emph{green} line) and in the radiation-pressure case (\emph{magenta}).
    \label{fig:dust_tau_profile}
    }
\end{figure}

In Fig.~\ref{fig:cloud_morphology} we provide an overview of the complex MC morphology at the reference evolutionary stage, by zooming on the central $20 \times 20$~pc$^2$ of the simulated box. At this stage, massive stars have already produced enough ionizing photons to form $\HII$ bubbles around them, and their feedback-driven turbulence led to the formation of filamentary structures of neutral gas.

The dust surface density (top-left panel) is distributed into several filaments with densities $\Sigma_{\rm dust} \gtrsim 10~\surfdpc$, in contrast with lower-density regions at $1~\surfdpc$; few clumps are also visible where filaments intersect each other.
The top-center panel shows instead the stellar mass distribution of the $729$ stars in place at this epoch, with a mass between $1-84~\msun$, and a total stellar mass of $M_\star = 2.4\times 10^3~\msun$; as expected, stars feature a more compact distribution when compared to the dust component.
The top-right panel illustrates the photon density-weighted Habing flux $G_0$ ($6 {\rm eV} < h\nu < 13.6 {\rm eV}$) computed on-the-fly in the \citetalias{Decataldo2020} simulation; the patchy distribution of the Habing flux reflects the impact of stellar radiation in their close surrounding, with high-mass stars imprinting their radiative feedback over larger distances, also because of stellar winds decreasing gas density in the immediate vicinity of stars. At this resolution, the $G_0$ flux spans $7$ orders of magnitude ($10^0-10^7$), with a skewed distribution toward large values, as seen by looking at the quantiles of the distribution, i.e. $\langle G_0 \rangle = 63.4_{-48.4}^{+381.4}$.
 
The bottom left and middle panels show the morphology of the MC emission in the UV ($0.1-0.3~\mum$) and IR ($8-1000~\mum$) bands. The UV emission toward the chosen line of sight shows a diffuse component at $S_{\rm UV} \sim (1-10) \times 10^{-4}~\surfl$, with few pixels with high values ($S_{\rm UV} \gtrsim 1~\surfl$) corresponding to the locations of unobscured luminous stars. Dust attenuation plays a major role in shaping the UV emission, since only $L_{\rm UV} \simeq 3.8 \times 10^5~\lsun$ is collected toward the chosen line of sight, corresponding to the $\simeq 13.1\%$ of the intrinsic luminosity of $L_{\rm UV, int} \simeq 2.9 \times 10^6~\lsun$. Most of this emission comes from stars whose stellar feedback cleared out the surrounding gas.
This trend is even more apparent by looking at the distribution of the UV optical depth $\tau_{\rm UV}$\footnote{The UV optical depth map is computed from $S_{\rm UV} = S_{\rm UV int} e^{-\tau_{\rm UV}}$ on a pixel-by-pixel base. Here $S_{\rm UV}$ and $S_{\rm UV int}$ are the UV surface brightness with and without dust attenuation respectively. The maps are smoothed out with a 2D Gaussian kernel with 1 pixel of standard deviation before the computation of the optical depth.}, which shows several peaks corresponding to dust-obscured stars embedded inside small high-density clumps or filaments. The optical depth distribution has a pixel-based median of $\tau_{\rm UV}=2.1^{+2.8}_{-1.6}$, thus most stars are optically thick in the UV band, with extreme peaks at $\tau_{\rm UV} > 10$.
The IR emission from dust thermal radiation features a spatial distribution that resembles the dust surface density distribution, with similar filamentary structures characterized by $S_{\rm IR} \sim 10^{-2}-10^{-1}~\surfl$. One highly emitting ($S_{\rm IR} \gtrsim 10~\surfl$) spot is also visible, corresponding to a dusty region irradiated by close stars, which end up being completely dust-obscured in the UV. The total IR luminosity is $L_{\rm IR} \simeq 3.0 \times 10^6~\lsun$, showing that most of the intrinsic radiation coming from the stars is re-emitted in the IR, due to the presence of dust giving a high $\tau_{\rm UV}$. 

We now discuss the cloud morphology in a more quantitative way. In the left panel of Fig.~\ref{fig:dust_mass_density_profile} we show the cumulative radial mass profile for the dust and stars (normalized to their respective total mass).
The stellar component is more compact than the dust component (see also Fig. \ref{fig:cloud_morphology}), with the $90\%$ of the total mass enclosed in a radius of $r_{*,90}\simeq 7.4$~pc and $r_{\rm dust, 90} \simeq 13$~pc respectively.
However the two profiles are quite similar and reach a plateau at the same radius, emphasizing that star formation follows the gas mass distribution.
As a reference, we also report the mass profile for the \citetalias{Sommovigo:2020} model in the uniform density case and in the radiation-pressure case. In the homogeneous case, the dust mass distribution is by construction $\propto r^3$, and it is less concentrated in the inner $\sim 10$~pc of the MC with respect to the \citetalias{Decataldo2020} simulations, however the radius containing $90\%$ of the total dust mass is comparable, with $r_{\rm dust, 90}^{\rm S20 \ hom} \simeq 15$~pc. In the radiation-pressure case, the central regions are almost deprived of their content as dust is pushed away. The mass profile recovers the shape of the homogeneous case only at $r\approx 7$~pc. However, the radius containing $90\%$ of the total mass is comparable, $r_{\rm dust, 90}^{\rm S20 \ RP} \simeq 15.3$~pc. The total dust mass in the radiation-pressure case is $1.7~\times 10^2~\msun$, as the $\HII$ region almost reaches the outer edge of the cloud.

We now analyze the absorption properties of the reference MC, by using the dust distribution to compute the cloud optical depth in different bands \footnote{We compute the optical depth from the dust distribution with the following method. We randomly select $100$ lines of sight (l.o.s.) from the center of the MC. We divide each l.o.s. in $100$ log-spaced radial bins from $0.05$~pc (comparable to $\Delta x$, i.e. the resolution from the \citetalias{Decataldo2020} simulation) up to $\sim 52$~pc (the diagonal of the box). For each l.o.s. a cylinder with radius $2\Delta x$ is considered to compute the dust surface density at each radial bin $\Sigma_{\rm dust}^{\rm bin}$, then the optical depth for the single radial bin is obtained via $\tau_\lambda^{\rm bin} = \Sigma_{\rm dust}^{\rm bin} \kappa_{\lambda}$, where $\kappa_\lambda$ is the mass absorption coefficient at wavelength $\lambda$. The contribution of each bin is then integrated from the center up to the limits of the box. The optical depth computed in this way represents the slab opacity, e.g. the opacity that a source placed in the center would experience along the specific l.o.s.}. In top panel of Fig.~\ref{fig:dust_tau_profile} we show the UV optical depth at $1600~\angstrom~$ ($\tau_{UV}$), integrated from the center of the cloud, as a function of the cloud radius; $\tau_{UV}$ has been computed for 100 lines of sight, and the median is shown as a solid line, while the shaded region shows its variance. The median UV optical depth rapidly increases from $\tau_{UV} \sim 10^{-2}$ to $\tau_{UV} \sim 1$ at $r \sim 1$~pc, and finally saturates at $\tau_{UV} \sim 30$ at $\sim 7.3$~pc, a smaller distance than the radius at which $90\%$ of the dust mass is enclosed, $r_{\rm dust, 90} \simeq 13.1$~pc. 
The variation of the optical depth between different lines of sight spans $0.5$~dex, which is relatively modest considering later stages of the evolution. 

We also compare the slab opacity discussed above with the effective opacity, defined as $\tau_{\lambda, \rm eff}$ such that $F_{\lambda} = F_{\lambda, 0} e^{-\tau_{\lambda, \rm eff}}$, where $F_{\lambda}$ is the detected flux and $F_{\lambda, 0}$ the flux that would be received without dust. This quantity provides an estimate of the radiation that is effectively removed from the line of sight considered, accounting for the relative distribution between stars and dust, plus radiative transfer effects. 
The effective UV opacity ranges between $\tau_{\rm UV, eff} = 2.0-4.6 $ for the six instruments used for the post-process RT probing different l.o.s., with the reference l.o.s. having $\tau_{\rm UV, eff} = 2.0$ (cfr. with the $\tau_{\rm UV}$ map in the bottom right panel in Fig.~\ref{fig:cloud_morphology}). This value implies an overall UV transmission along this line of sight of $e^{-\tau_{\rm UV, eff}} \approx 0.14$, consistent with the UV map in Fig.~\ref{fig:cloud_morphology}.

We also compare the optical depth of the simulated MC with the analytical model by \citetalias{Sommovigo:2020}, for the homogeneous case and the radiation-pressure case. In the first case, it increases from $\tau_{\rm UV} = 0.1$ at $r \approx 0.15$~pc to $\tau_{\rm UV} \sim 10$ at $ r \approx 15$~pc, as in this model $\tau_{\rm UV}^{\rm S20 hom} \propto r$. 
Despite the cloud considered in the analytical model having the same mass as the cloud in the simulation, the median optical depth in the simulated MC has a larger value at the cloud boundary with respect to the analytic model, because most lines of sight pass through high-density clumps.
In the radiation-pressure model, the inner region of the cloud is significantly dust-deprived by the effect of radiation pressure (cfr. with Fig.~\ref{fig:dust_mass_density_profile}). As a result, the UV optical depth remains very low and becomes $\tau_{\rm UV} \approx 1$, only at $r\approx 6$~pc. At the cloud boundary it reaches $\tau_{\rm UV}^{\rm S20 \ RP} \sim 4$, thus even in this model the cloud is overall UV optically-thick. This high value of $\tau_{\rm UV}$ is consistent with the fact that the dust mass retained by the cloud in this model is $\approx 73\%$ of the original one. 
From this comparison we note that, despite their simple geometric assumptions, both analytical models have some merit in describing the optical depth in the simulated MC. In particular, the RP model provides the best estimate of the effective UV optical depth at the edge of the cloud, whereas the homogeneous model better reproduces the profile of the physical UV optical depth.

In the bottom panel of Fig.~\ref{fig:dust_tau_profile} we instead show the IR optical depth. Despite the high-density filaments and clumps, the simulated MC is always optically thin in the IR among all the considered lines of sight, due to the fact that the dust opacity $\kappa_{100~\mum}$ is a factor of $\approx 2000$ lower than $\kappa_{1600~\angstrom}$. The IR optical depth has a median value increasing from $\tau_{IR} \sim 10^{-5}$ to $\tau_{IR} \sim 1.6 \times 10^{-2}$ at the saturation distance of $\sim 5$~pc. Even considering the variance between different lines of sight, the IR optical depth stays always below $0.1$. 
Similarly, for the homogeneous model in \citetalias{Sommovigo:2020}, the IR optical depth increases from $\tau_{IR}^{\rm S20 \ hom} \sim 4.8 \times 10^{-3}$ to $\tau_{IR}^{\rm S20 \ hom} \sim 4.8 \times 10^{-3}$.
Instead, in the radiation-pressure model, the IR optical depth is much lower, analogously to the UV case. At the cloud boundary, $\tau_{IR}^{\rm S20 \ RP} \sim 2.2\times 10^{-3}$.

\subsection{Cloud emission} \label{sec:cloud_em}

\begin{figure}
    \centering
    \includegraphics[width=0.49\textwidth]{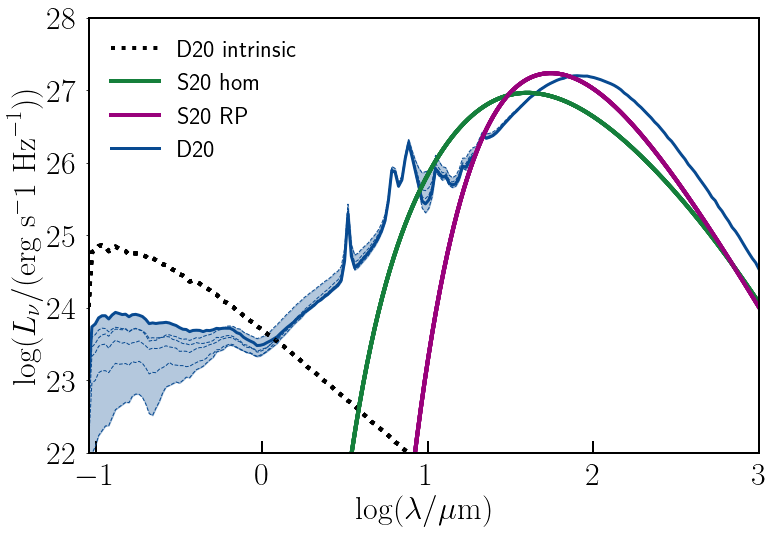}
    \caption{
    Cloud Spectral Energy Distribution (SED, $L\nu$) for the reference MC as a function of wavelength ($\lambda$). 
    The solid \emph{blue} line shows the UV-to-FIR emission of the cloud obtained from the RT post-process along the reference line of sight. The dashed \emph{blue} lines correspond to the other five l.o.s. The shaded area brackets the fluxes of all the l.o.s. adopted.
    The dashed \emph{black} line indicates the intrinsic (dust-transparent) emission of the cloud along the same line of sight.
    As a reference we also show the emission of a local cloud from the analytic model by \citetalias{Sommovigo:2020} in the homogeneous case (\emph{green} line) and in the radiation-pressure case (\emph{magenta} line). 
    \label{fig:cloud_SED}}
\end{figure}

In Fig.~\ref{fig:cloud_SED} we analyze the emission of the reference MC, by comparing the intrinsic (e.g. dust-transparent) SED and the SED observed by the different instruments after the RT post-process. The intrinsic emission at $0.1~\mum \lesssim \lambda \lesssim 1~\mum$ is attenuated up to one order of magnitude for the reference l.o.s., which is the least-attenuated l.o.s. The observed UV emission has a scatter of one order of magnitude, as also suggested by the estimate of the effective UV optical depth shown in Fig.~\ref{fig:dust_tau_profile}. 
Most of the l.o.s. show a flat slope in the UV emission, however the most attenuated ones show an increase toward longer wavelengths.
The mid-infrared (MIR) part of the SED is mainly characterized by emission from PAHs at $3.3$, $6.2~\mum$ and $11.2~\mum$, and by the silicate absorption feature at $9.7~\mum$.
At longer wavelengths, the SED shows the characteristic IR bump corresponding to dust thermal emission, with a peak at at $\lambda_{\rm peak}=80~\mum$.  

As a reference, we also show the results from the analytic model by \citetalias{Sommovigo:2020}. In the homogeneous case, the IR emission peaks at a shorter wavelength $\lambda_{\rm peak}=40~\mum$ with respect to the simulated MC. The peak of the IR emission is related with the dust temperature distribution in the cloud. Hence, the difference in the peak wavelength suggests that the dust component in \citetalias{Sommovigo:2020} settles on higher temperatures than the simulated reference MC. This is due, in the homogeneous analytical case, to the presence of dust in the central region of the cloud, directly irradiated by stars without experiencing any shield effect. We further discuss this point in Sec.~\ref{sec:dust_temp}. 
In the case with radiation pressure, the IR emission peaks at a longer wavelength $\lambda_{\rm peak}=55~\mum$ with respect to the homogeneous model. This is due to the fact that in this model most of the dust is at large distance from the center, thus the flux contributing to the dust heating is lower because of geometrical dilution of the flux. As a result, the temperature settles to lower values with respect to the homogeneous model (see Sec.~\ref{sec:dust_temp}). Moreover, when dust mass is pushed outside and accumulated in an outer shell with a larger volume, the density becomes lower. However, the dust UV optical depth at the cloud boundary is still larger than 1 ($\tau_{\rm UV}^{\rm S20 \ RP} \sim 3$, cfr. with Fig.~\ref{fig:dust_tau_profile}), therefore the overall IR emission is the same as in the homogeneous model, because of the LTE assumption in the analytic model (e.g. the IR emission is equal to the absorbed UV luminosity).

\subsection{Dust temperature in the cloud} \label{sec:dust_temp}

\begin{figure}
    \centering
    \includegraphics[width=0.49\textwidth]{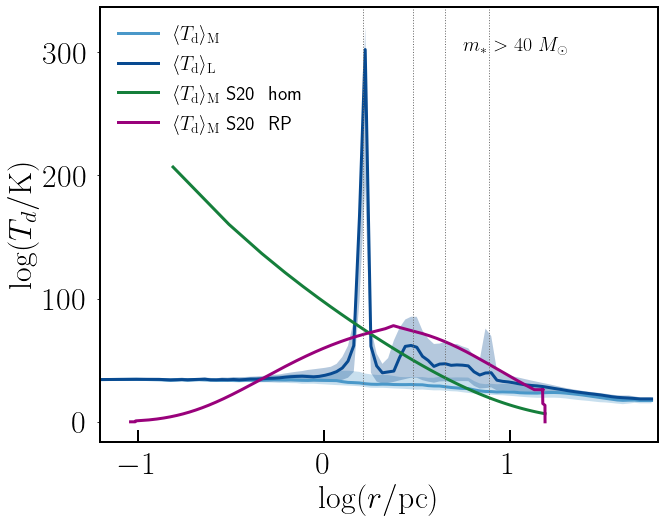}
    \hfill
    \caption{
    Dust temperature radial profile of the reference MC. 
    The \emph{light blue} (\emph{dark blue}) line shows the median of the mass-weighted (luminosity-weighted) dust temperatures in spherical shells; the corresponding shaded areas bracket the $16$th-$84$th percentiles.
    Dotted vertical lines mark the positions of massive ($m_\star\geq 40~\msun$) stars.
    As a reference, we show the temperature profile for the \citetalias{Sommovigo:2020} model in the homogeneous case (\emph{green} line) and in the radiation pressure case (\emph{magenta} line).
    \label{fig:dust_temperature_profile}
    }
\end{figure}

\begin{figure*}
    \centering
    \includegraphics[width=0.9\textwidth]{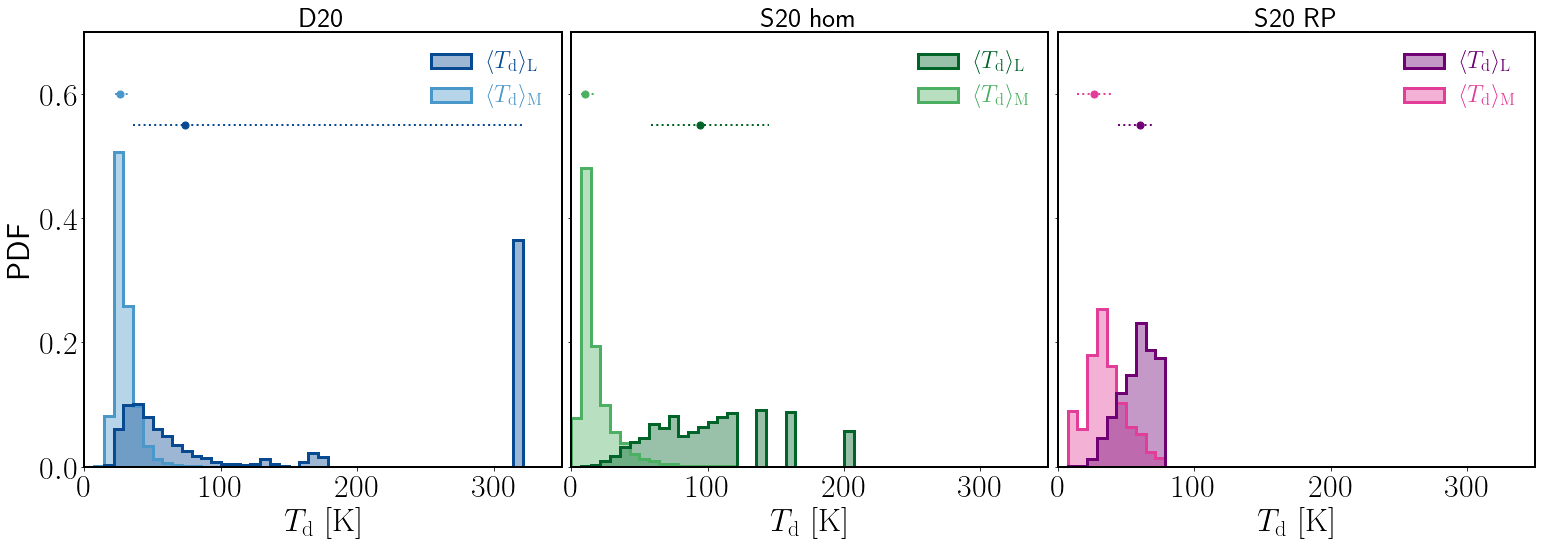}
    \hfill
    \caption{
    Dust temperature distribution in the reference MC. 
    Probability distribution functions (PDF) weighted by mass and luminosity, with $\Delta T\approx 7.1$~K and $50$ bins.
    \textbf{Left panel} refers to the simulated MC in \citetalias{Decataldo2020}, \textbf{middle panel} to the homogeneous model in \citetalias{Sommovigo:2020} and \textbf{right panel} to the radiation-pressure model.
    Circles indicate the median of each distribution, whereas the dotted lines bracket the $16$~th and $84$~th percentiles.
    \label{fig:Tdust_PDF}
    }
\end{figure*}

The IR emission of the MC discussed in Sec.~\ref{sec:cloud_em} originates from thermal radiation by dust grains, whose temperature depends on the luminosity of the radiation sources and the distance from them. 
In Fig.~\ref{fig:dust_temperature_profile} we show the dust temperature radial profile of the reference MC. In particular, we consider for each spherical shell the \emph{mass-weighted} dust temperature \TdMW{} and the \emph{luminosity-weighted} dust temperature \TdLW{}\footnote{For the luminosity-weighted temperature, we assign a luminosity $L_{\rm cell} \propto m_{\rm cell} T_{\rm cell}^{4+\beta}$ to each cell with mass $m_{\rm cell}$ and temperature $T_{\rm cell}$. We fix the dust emissivity index $\beta=2$ for this computation, consistent with the far-IR opacity of the \citep{Weingartner:2001} model.}. The first quantity provides an estimate of the dust temperature of the bulk of the dust component, whereas the second is biased toward the brightest dust grains, likely the one strongly irradiated by close stars. 
The median of \TdMW{} is $\approx 34$~K in the innermost region ($r < 1$~pc) and slowly decreases down to $17$~K at the cloud boundary. It shows a very small scatter, suggesting that most of the dust mass spans a limited range in dust temperatures. 
The median of \TdLW{} follows \TdMW{} in the innermost region. However, it features few strong peaks up to $\approx 170$~K, with some particularly prominent at $1.6$~pc, $3$~pc, $4.5$~pc and $7.8$~pc. These peaks correspond to spherical shells containing massive stars ($m_\star > 40~\msun$) with a strong radiation field. The position of such massive stars is also emphasized in the figure and it is remarkably consistent with the peak in the luminosity-weighted profile. Moreover, the smaller the radius, the higher and narrower are the \TdLW{} peaks, because of the spherical averaging.

For comparison, we also show the dust temperature from the analytical model \citetalias{Sommovigo:2020}, averaged over the mass of dust grains. In the homogeneous model, the dust temperature peaks at $\approx 200$~K in the center and drops toward larger distances. This behaviour is expected because in this model, all the stars are placed in the center of the cloud, therefore the dust in the innermost region is heated toward high temperatures, whereas at higher distances the flux is reduced because of geometrical dilution and dust attenuation. 
The high-temperature component in the \citetalias{Sommovigo:2020} homogeneous model is responsible for an IR SED which peaks toward shorter wavelength with respect to the simulated MC (cfr with Fig.~\ref{fig:cloud_SED}). 
Instead, in the radiation-pressure case, the dust temperature shows a very different profile, due to the different spatial distribution. It has a very low values in the central region and then increases going outward as the optical depth increases. It reaches a peak of $T_{\rm d} \approx 80$~K at $r \approx 2$~pc, and then slowly decreases again as the outer shells experience an higher UV optical depth and therefore absorb less flux from the central stars.

We further investigate the dust temperature within the MC by considering the overall temperature distribution instead of the spatial distribution: we show in Fig.~\ref{fig:Tdust_PDF} the Probability Distribution functions (PDF) weighted by mass (\emph{red}) and luminosity (\emph{blue}) over all the dust cells in the computational box of the RT post-process. The mass-weighted PDF has a narrow distribution, with a median at ${\langle T_{\rm d} \rangle}_{\rm M} = 25.9^{+7.0}_{-3.8}$~K, where the error bars mark the $16$th and $84$th percentiles. This estimate indicates a mild skeweness toward higher temperatures, due to the presence of the hottest dust cells, which constitutes only a small fraction of the total dust mass. As a reference, the total dust mass above $80$~K is $\approx 0.14~\msun$, e.g. $\approx 0.02\%$ of the total dust mass (cfr. also with the right panel in Fig.~\ref{fig:Tdust_evolution}).
The hot dust component is well emphasized by the luminosity-weighted PDF, which shows a wide distribution, with multiple peaks at $T>100$~K. The median of this distribution is: ${\langle T_{\rm d} \rangle}_{\rm L} = 74.0^{+247}_{-38.5}$~K. 

In the homogeneous model of \citetalias{Sommovigo:2020}, most of the dust tends to be colder than in the simulated MC, but also in this case a mild skeweness toward higher temperature is present, with ${\langle T_{\rm d} \rangle}_{\rm M} = 9.8^{+7.6}_{-2.5}$~K. This asymmetry in the distribution is driven by the central warm dust regions. The total dust mass above $80$~K is $0.46~\msun$, a factor of $3$ higher than in the simulated MC. As a result, also the median of the luminosity-weighted distribution is higher, ${\langle T_{\rm d} \rangle}_{\rm L} = 94.2^{+50.6}_{-35.9}$~K. This is reflected into the FIR emission, which peaks at shorter wavelengths with respect to the simulated MC (cfr with Section.~\ref{sec:cloud_em} and Fig.~\ref{fig:cloud_SED}). However, the luminosity-weighted temperature distribution is more compact than in the simulated MC, as there are no extreme regions with dust temperature $T>300$~K due to the presence of small pockets of dust very close to a luminous star.

In the the radiation-pressure model of \citetalias{Sommovigo:2020} the dust temperature covers a much narrower range, because it is pushed at large distance from the center. As a result, the mass-weighted and luminosity-weighted distribution are much similar, with ${\langle T_{\rm d} \rangle}_{\rm M} = 26.1^{+14.4}_{-11.9}$~K and
${\langle T_{\rm d} \rangle}_{\rm L} = 59.8^{+11.6}_{-15.9}$~K respectively.

\section{Evolution of MC properties} \label{sec:evolution}

Having thoughtfully analyzed the reference MC, which has been selected at $t \approx 1.1$~Myr since the beginning of the star formation, we can now explore the full time evolution. In particular, we investigate the evolution of its emission and absorption properties.

\subsection{Dust temperature evolution}

\begin{figure*}
    \centering
    \includegraphics[width=0.49\textwidth]{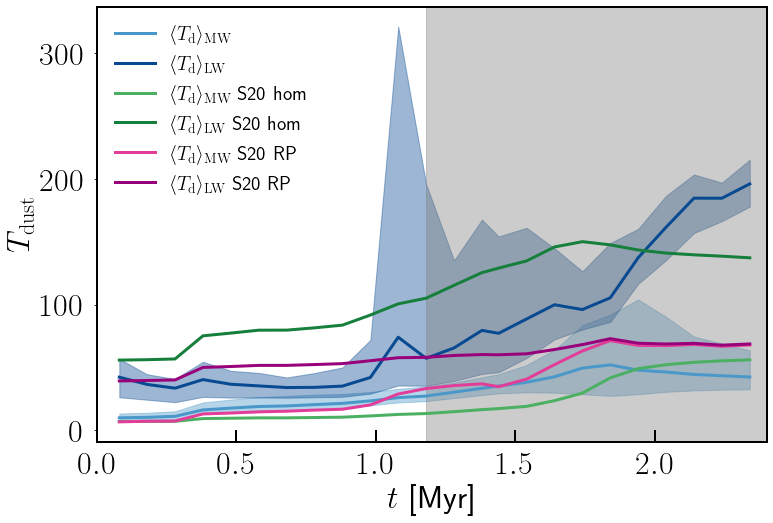}
    \hfill
    \includegraphics[width=0.49\textwidth]{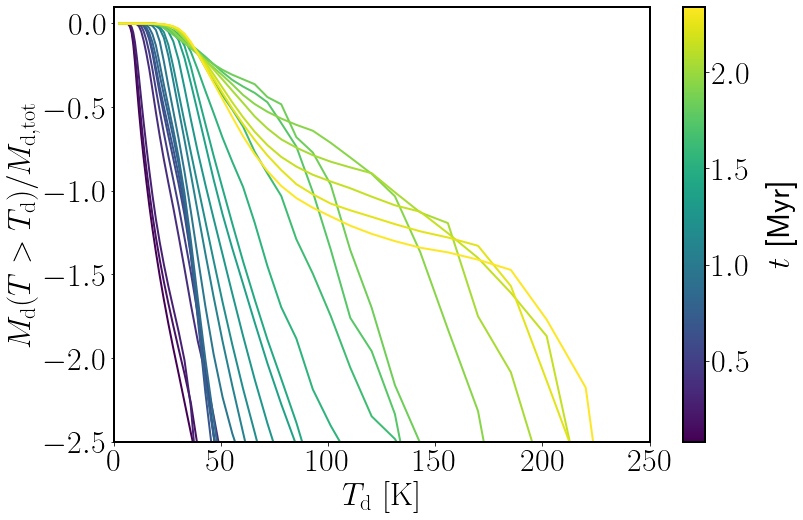}
    \caption{
    Evolution of the dust temperature during the cloud lifetime, after the onset of star formation. 
    \textbf{Left panel.} The \emph{light blue} (\emph{dark blue}) solid line shows the median of the mass-weighted (luminosity-weighted) distributions for the simulated MC. The shaded areas mark the $16$-th and $84$-th percentiles of each distribution. The \emph{light green} (\emph{dark green}) line shows the evolution of the mass-weighted (luminosity-weighted) distributions for the local cloud model by \citetalias{Sommovigo:2020} in the uniform case. For the radiation pressure case, we show in \emph{light magenta} (\emph{dark magenta}) the mass-weighted (luminosity-weighted) distributions.
    The shaded area highlights the period when the cloud is dispersed according to the radiation-pressure model from \citetalias{Sommovigo:2020} (cfr. with Sec.~\ref{sec:S20_model}.) 
    \textbf{Right panel.} Cumulative distribution of the fractional mass of dust ($M_{\rm d}(>T_{\rm d})/M_{\rm d, tot}$) that has a temperature larger than $T_{\rm d}$, colour-coded according to the evolution time $t$ of the MC.
    \label{fig:Tdust_evolution}
    }
\end{figure*}

We now examine how the evolution of the radiation field and of the dust morphology affects the dust temperature distribution throughout the cloud lifetime.

We use the median values of the mass-weighted and luminosity-weighted distributions (\TdMW{} and \TdLW{} respectively) and their variance as indicators of the dust temperature distributions (cfr. with Fig.~\ref{fig:Tdust_PDF} relative to the fiducial MC). In Fig.~\ref{fig:Tdust_evolution} we plot the median values of the mass-weighted and luminosity-weighted distributions for the simulated MC and the local cloud analytical models by \citetalias{Sommovigo:2020}. 
We find that \TdMW{} and \TdLW{} remain almost constant at $\approx 20-30$~K and $40-50$~K respectively during the first $1$~Myr of the cloud evolution, when the star formation proceeds at a low rate ($\SFR \lesssim 10^{-2}~\msunyr$) and the number of very bright stars is low. 
This is further emphasized by the evolution of the dust mass fraction above a certain temperature $T_{\rm d}$ at different stages of the cloud evolution, shown in the right panel of Fig.~\ref{fig:Tdust_evolution}. At this stage, less than $1\%$ of the total dust mass is heated above $50$~K.  

After $t\simeq 1$~Myr, the radiation field increases in intensity and both \TdMW{} and \TdLW{} start to increase, but their evolution is quite different. The former, tracing the bulk of the dust mass, which is mostly heated by the diffuse radiation, grows mildly up to $\approx 52$~K at $t\simeq 1.8$~Myr, and then slowly decreases in the later stages of the cloud evolution, when the star formation rate in the cloud drops because of the cloud photo-ionization. Thus, \TdMW{} mostly follows the evolution of the star formation rate. Instead, \TdLW{} grows from $t\simeq 1$~Myr up to the end of the simulation, reaching $\approx 200$~K. This is because the dusty clumps close to bright stars, which are the ones mainly traced by \TdLW{}, keep receiving photons by these already formed stars, despite the overall star formation in the cloud declines. 
The importance of small, hot clumps in driving the evolution of \TdLW{} can be understood in terms of the cumulative mass fraction above a given temperature. Between $1.0 \lesssim t \lesssim 1.8$~Myr, the fraction of mass above $100$~K progressively increases up to $\approx 25\%$ of the total. After $1.8$~Myr, it decreases down to less than $10\%$, however this lower amount of dust above $100$~K is heated to higher temperatures. This can be appreciated by the tail in the profile, with an increasing fraction of dust above $150$~K as the cloud evolves.

We compare these results from the RT calculations with the predictions from the analytical model by \citetalias{Sommovigo:2020} for a local cloud. In the uniform case, \TdMWS20{} remains constant at $10-20$~K up to $t\simeq 1.8$~Myr, as most of the radiation from the central stars is absorbed by the innermost shells in the cloud and therefore the bulk of the dust mass does not change significantly its temperature. After the peak in the SFR, with the cloud having less dust mass (and thus less attenuation from the inner shells), \TdMWS20{} quickly rises up to $\approx 55$~K. On the other hand, \TdLWS20{} starts from $\approx 56$~K at the beginning of the cloud life, and moderately rises up to $\approx 150$~K at $t\simeq 1.8$~Myr, as the inner shells experiences more and more radiation. When the star formation rate begins to drop, \TdLWS20{} follows too, reaching $\approx 137$~K at the end of the simulation.
In the radiation-pressure case, the \TdMWS20{} evolves similarly to the simulated \TdMW{} up to $t\simeq 1.5$~Myr; while the radial profiles appear quite different at $r \lesssim 10$~pc (see Fig.~\ref{fig:dust_temperature_profile}), this does not affect significantly the average temperature, since only $\approx 10\%$ of the dust mass is located within this radius (see Fig.~\ref{fig:dust_mass_density_profile}). 

From $t\gtrsim 1.5$~Myr, the dust mass loss due to radiation pressure becomes relevant in the analytic model, thus the dust mass retained in the cloud gets heated to higher temperature at the same radiation field. As a result, \TdMWS20{} increases to $\approx 70$~K and remains constant after $t\gtrsim 1.8$~Myr.
On the other hand, \TdLWS20{} presents a very mild evolution throughout the cloud lifetime, starting at $40-50$~K and slowly growing up to $\approx 70$~K at $t\simeq 1.8$~Myr. This is a consequence of the fact that radiation pressure pushes the dust away from the central stars, so that there are no dust shells that experience a strong radiation field. This is as also shown in the detailed spatial distribution (cfr. with Fig.~\ref{fig:dust_temperature_profile}) and in the dust temperature distribution (Fig.~\ref{fig:Tdust_PDF}, right panel). At $t \gtrsim 1.8$~Myr \TdLWS20{} remains constant at almost the same value as \TdMWS20{}, because at this stage the intense radiation pressure has confined most of the dust mass in the outermost shells, which are therefore heated at the very similar temperatures.

This comparison shows that the radiation-pressure model is able to provide a realistic description of the bulk of the dust temperature within the cloud and of its infrared emission (cfr. with Fig.~\ref{fig:cloud_SED}), until $t \simeq 1.5$~Myr, i.e. around $0.3$~Myr after the moment the cloud is effectively dispersed in this model. However, due to the simplistic stellar-to-dust geometry, it is not able to capture the high-temperature regions, corresponding to clumps irradiated by close bright stars.

\subsection{Evolution of the emission properties}

\begin{figure}
    \centering
    \centering\includegraphics[width=0.49\textwidth]{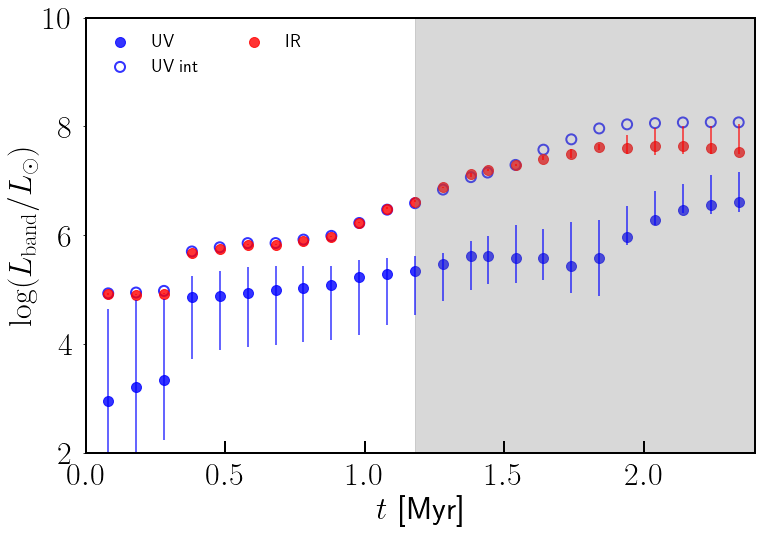}
    \caption{
    Luminosity evolution for the MC.
    Total UV (\emph{blue}) and IR (\emph{red}) luminosity as a function of cloud age since the onset of star formation. Empty circles show the intrinsic stellar radiation, i.e. ignoring dust absorption. Full circles indicate the median luminosity among the the different l.o.s. considered, whereas the error bars bracket the minimum and maximum luminosity. 
    The shaded area highlights the period when the cloud is dispersed according to the radiation-pressure model from \citetalias{Sommovigo:2020} (cfr. with Sec.~\ref{sec:S20_model}.) 
    \label{fig:lum_history}
    }
\end{figure}

\begin{figure}
    \centering
    \includegraphics[width=0.49\textwidth]{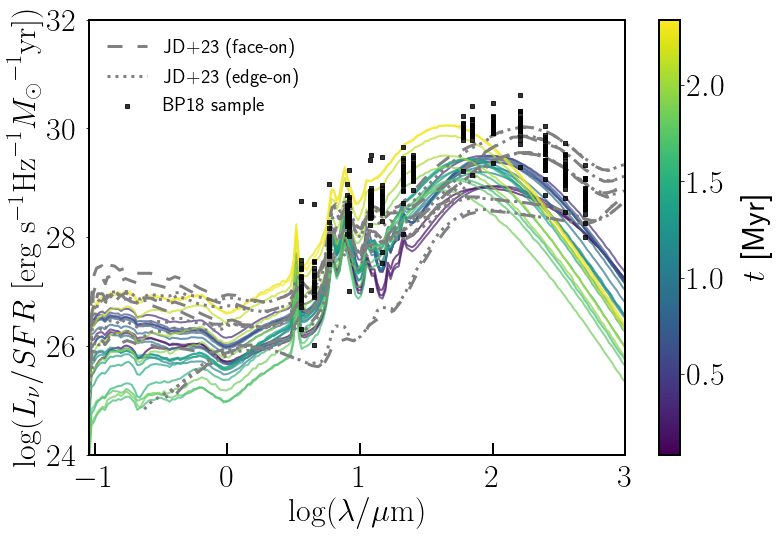}
    \caption{
    Evolution of the MC SED as a function of wavelength, normalized with the SFR.
    Each line corresponds to a different evolution time $t$ of the MC, as shown in the colorbar. For the sake of clarity, at each $t$ we show only the SED for the reference line of sight.
    Dashed grey lines show the results of the RT simulations by \citetalias{Jaquez-Dominguez2023} for their snapshots from $0.8$~Myr to $7.5$~Myr for their face-on instrument (dashed lines) and edge-on instrument (dotted lines).
    Black points indicate the photometric data relative to the $28$ active clouds analyzed in \citet{BinderPovich2018ApJ...864..136B}.
    \label{fig:cloud_SED_evolution}
    }
\end{figure}

In Fig.~\ref{fig:lum_history} we show the total radiation emitted from the cloud in the UV ($0.1-0.3~\mum$) and IR ($8-1000~\mum$) bands for the selected snapshots at all the simulated lines of sight, emphasizing the median values.
The UV radiation produced by stars (empty blue circles) increases throughout the MC evolution loosely following the average trend of the star formation history (cfr. with Fig.~\ref{fig:sfr_history}). At $t \lesssim 0.3$~Myr, the intrinsic emission is $\sim 10^5~\lsun$, of which only $\approx 1\%$ survives dust attenuation along the median l.o.s.
At this stage, the cloud UV emission also shows a strong dependence with the l.o.s., with variation of almost three orders of magnitudes; this behaviour is caused by the fact that at this stage only few stars ($\approx 70$, with a total mass of $M_\star \approx 170~\msun$) are in place, making geometric effects very important for the dust attenuation.
At $t \sim 0.3$~Myr, the cloud emission increases of around one order of magnitude both in the UV and IR. This is due to the birth of a star with $M_\star > 40~\msun$, which significantly boosts the photon production ($L_\star \propto m_\star^3$). After this episode, the overall luminosity remains broadly constant in both the UV and IR bands up to $t\sim 1$~Myr, and also the dependence of the UV emission with the l.o.s. is significantly reduced. Moreover, during this stage up to $t\sim 1.5$~Myr, the IR emission is almost equal to the intrinsic UV emission, e.g. $L_{\rm UV} \simeq L_{\rm IR}$.
After this point, the intrinsic UV emission increases again, reaching $3.7\times 10^7~\lsun$ at $t\simeq 1.7$~Myr, when the cloud begins to photo-evaporate. However, during this period, the dust-attenuated UV luminosity does not increase with the same pace, and actually decreases after $t\simeq 1.4$~Myr, suggesting an increase in the effective optical depth for the whole MC. Between $1.3 \lesssim t \lesssim 1.5$~Myr, the MC showcases a slightly higher IR luminosity with respect to the intrinsic UV luminosity. This is a consequence of the increased intrinsic emission in the range $0.3-1.0~\mum$, whose contribution to the dust heating becomes significant.
In the latest stages ($t>1.7$~Myr), the SFR in the simulated MC flattens and then starts to drop, however the UV intrinsic stays constant, because massive stars are still seeded. Instead, the UV emission steadily increases among all lines of sight, as a result of an increased UV escape fraction, due to the cloud photo-evaporation (see later on in Fig.~\ref{fig:UV_escape}).
We also notice that during this epoch the IR emission is a factor of $\approx 2$ lower than the UV intrinsic luminosity, whereas they were almost comparable before $t \sim 1.5$~Myr. This is due by the scattered light outside the l.o.s., as can be seen by the errorbars of the UV and FIR in Fig.~\ref{fig:lum_history}.

We further study the evolution of the emission by considering the whole SED. In Fig.~\ref{fig:cloud_SED_evolution} we show the SED of the MC at each of the selected snapshots, color-coded based on the time after the beginning of the star formation. Overall, the SEDs show a similar shape throughout the evolution, with the total luminosity that increases as the cloud evolves. However, by inspecting closely each band, few significant trends emerge.
The UV emission increases up to $t\simeq 1.5$~Myr, then decreases for a short period up to $t \simeq 1.8$~Myr, and then quickly rises again, as in Fig.~\ref{fig:lum_history}. 
The UV SED presents a large variety of slopes\footnote{We define the UV slope $\beta_{\rm UV}$ in the wavelength range $1600-2500~\angstrom$ as $\beta_{\rm UV} = \log(L_{\lambda_1}/L_{\lambda_2})/\log(\lambda_1/\lambda_2)$, where $(\lambda_1,\lambda_2) = (1600, 2500)~\angstrom$~. \label{foot_betaUV}}.
At $t\lsim 0.3$~Myr, the MC shows positive slopes, $\beta_{\rm UV} \simeq 0.5$; at $0.3 \lsim t \lsim 1.2$~Myr it is mainly characterized by mild negative slopes $\beta_{\rm UV} \simeq -0.5$; after $t\lsim 1.2$~Myr, the slopes tend to flatten up to $t\simeq 1.8$~Myr, when they tend to decrease again down to $\beta_{\rm UV} \simeq -0.5$. 
In the range $1~\mum \lesssim \lambda \lesssim 10~\mum$, the shape of the SED is remarkably similar at each time frame and simply changes in magnitude. They all show the PAHs features at $3.3$ and $6.2~\mum$, and the silicate feature at $9.7~\mum$. The only exception to this trend is at $t \lesssim 0.3$~Myr, when the SEDs show a flatter slope in this wavelength range. At $\lambda \gtrsim 10~\mum$, the PAH features become less and less prominent as the MC evolves, because the dust temperature distribution shifts toward higher values (see Fig.~\ref{fig:Tdust_evolution}), raising the underlying continuum flux.  

Finally, the peak of FIR emission increases in magnitude by $\sim 3$ orders of magnitude throughout the MC evolution, but the wavelength of the peak also shifts toward shorter wavelengths, from $110~\mum$ at $t \approx 0.3$~Myr to $31~\mum$ at $t \approx 2.3$~Myr. This is due to the shift of the dust temperature distribution toward higher values as the MC evolves, because the intensity of the radiation field increases as more and more stars are seeded (cfr with the intrinsic UV emission in Fig.~\ref{fig:lum_history}).

We compare our results with the RT simulations by \citetalias{Jaquez-Dominguez2023}, plotting the SEDs relative to their snapshots from $0.8$~Myr to $7.5$~Myr, for their face-on and edge-on instruments. 
Our simulated SEDs at $t \lesssim 1$~Myr tend to have lower luminosities (normalized with the star formation rate) in the UV band with respect to the ones by \citetalias{Jaquez-Dominguez2023}, with the exception of their snapshots at $t=0.8$~Myr and $t=1.6$~Myr. At $t \gtrsim 1$~Myr the simulated SEDs between the two works span a comparable range in SFR-normalized luminosity. 
In the mid-infrared range the SEDs from both simulations are fairly consistent. Instead, in the FIR band, the SEDs by \citetalias{Jaquez-Dominguez2023} peak at longer wavelengths with respect to our simulated SEDs. This is due to the fact that the dust component settles to lower temperatures in \citetalias{Jaquez-Dominguez2023} with respect to our MC. We attribute this difference to the higher star formation efficiency in the simulation considered in this work. In fact, at $t \gtrsim 1$~Myr our simulated MC has a SFR $1-2$ orders of magnitude higher with respect to the typical values in the simulation from \citetalias{Jaquez-Dominguez2023}. As a result, the $M_*/M_{\rm gas}$ ratio is at least a factor $4-5$ higher throughout the evolution of the cloud. The lower dust-to-stellar ratio results in higher dust temperature in our simulated cloud.

We also compare our results with the photometric data of $28$ massive star-forming HII regions in the Milky-Way by \citet{BinderPovich2018ApJ...864..136B}. These clouds have an infrared luminosity between $4\times 10^4~\lsun$ and $2.3\times 10^7~\lsun$ and they have a SFR (inferred from the TIR luminosity) spanning the range $1.8\times 10^{-5}~\msunyr \lesssim \SFR \lesssim 10^{-2}\msunyr$. Therefore, they can be compared to our simulated SEDs up to the reference MC at $t \approx 1.1$~Myr.
The observed HII regions are quite consistent with the MIR flux from our simulations. Instead, the FIR data shows a FIR peak at longer wavelengths, suggesting that the simulations are missing a lower dust temperature component. If considering the overall IR emission, the disagreement with observations is milder. In fact, the simulated MC has a TIR luminosity of $L_{\rm TIR} \approx 10^6 - 3 \times 10^7$ for most of its lifetime, and the majority ($17$ out $28$) of the clouds in \citet{BinderPovich2018ApJ...864..136B} have $L_{\rm TIR} \approx 10^6 - 2 \times 10^7$.
We attribute the discrepancy between our synthetic SEDs and observations of local HII regions to the star formation in our simulated MC. In other words, the intense star formation (and feedback) in the MC from \citetalias{Decataldo2020} depletes the gas and therefore the dust component, which is then heated by UV radiation at temperatures which are higher with respect to the observations from the \citet{BinderPovich2018ApJ...864..136B} sample. This effect might be due to the absence of magnetic fields and to the lack of full turbulence driving in the \citetalias{Decataldo2020} setup. Indeed, both factors can increase the gas pressure support preventing gravitational collapse, in turn reducing the star formation rate efficiency and thus the depletion.

\subsection{Evolution of dust attenuation properties}

\begin{figure*}
    \centering
    \includegraphics[width=0.49\textwidth]{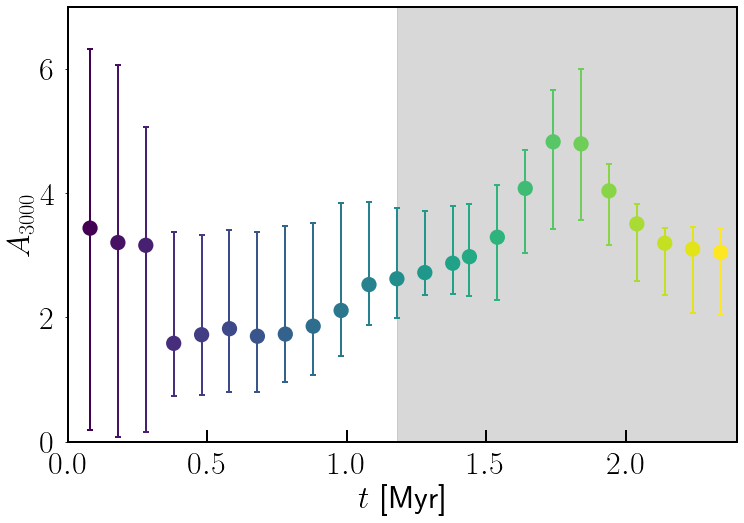}
    \hfill
    \includegraphics[width=0.49\textwidth]{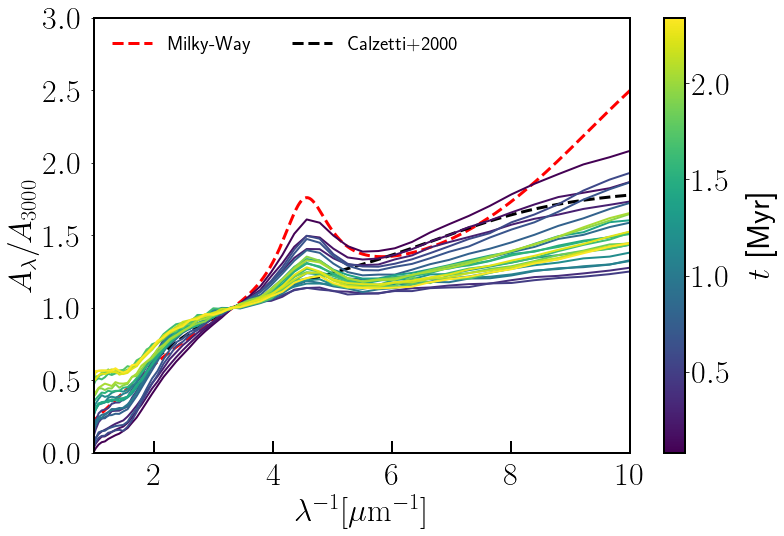}
    \caption{
    Evolution of the MC attenuation. 
    \textbf{Left panel:} Flux attenuation at $3000~\angstrom~$ at different times during the cloud evolution. The circles indicate the median $A_{3000}$ among the considered l.o.s., whereas the errorbars bracket the minimum and maximum values. 
    The shaded area highlights the period when the cloud is dispersed according to the radiation-pressure model from \citetalias{Sommovigo:2020} (cfr. with Sec.~\ref{sec:S20_model}.) 
    \textbf{Right panel:} flux attenuation (normalized to the attenuation at $3000~\angstrom~$) as a function of the inverse of the wavelength for the post-processed snapshots. The attenuation curves are shown only for the median line of sight. Dashed lines show the attenuation curve for the Milky-Way (\emph{red}) and the \citet{Calzetti:2000} one (\emph{black}).
    \label{fig:cloud_Alambda_evolution}
    }
\end{figure*}

In Fig.~\ref{fig:cloud_Alambda_evolution} we show the attenuation properties of the MC throughout its evolution. 
We first focus on a single wavelength, by studying $A_{3000}$, the attenuation at $3000~\angstrom~$, which is often used to pinpoint the attenuation curves.
The left panel of Fig.~\ref{fig:cloud_Alambda_evolution} shows $A_{3000}$ as a function of the cloud age. At the beginning, the median attenuation is $A_{3000} \approx 4$, with a large scatter among different l.o.s. Then, at $t\sim 0.4$~Myr, it drops to $A_{3000} \approx 2$ and the scatter between different lines of sight is progressively reduced as the cloud evolves and the number of stars formed increases.
From $t\sim 0.4$~Myr onwards, the median attenuation keeps increasing during the cloud evolution up to $A_{3000} \approx 5$ at $t \sim 1.8$~Myr (when the volume occupied by HI reaches a peak, cfr. with Fig.~9 in \citetalias{Decataldo2020}). After this period, the cumulative effect of stellar feedback leads to the cloud photo-evaporation, with the formation of many cavities and a dust loss of a factor $2$ since the beginning of star formation. As a result, $A_{3000}$ drops steadily down to $\approx 3$ at the end of the simulation.

In the right panel of Fig.~\ref{fig:cloud_Alambda_evolution}, we show the median attenuation curve among the different lines of sight, normalized to $A_{3000}$, at the same snapshots as in the left panel.
We notice that immediately after the onset of the star formation the attenuation curve is close to the Milky-Way attenuation curve, with a slightly less pronounced bump at $2175~\angstrom~$, and a less steep rise at short wavelengths. In the time period between $0.4 \lesssim t \lesssim 1$~Myr, the attenuation curve experiences a rapid variation: the bump is suppressed for a short time, then it reappears almost as strong as at the beginning and then it is suppressed again. At short wavelengths the curve remains flatter than the original Milkly-Way attenuation and becomes closer to the Calzetti curve \citep{Calzetti:2000}. 
After $t\sim 1$~Myr, the bump strength slowly increases, with some variation at different times, whereas the slope of the curve becomes flatter than the Calzetti curve.

We do not find a clear correlation between the strength of the $2175~\angstrom~$-bump and the attenuation $A_{3000}$, nor between the steepness of the attenuation curve and $A_{3000}$. We instead find a mild correlation with the relative variation of $A_{3000}$ across the lines of sight, defined as $\sigma_{A_{3000}} = A_{3000, \rm max} - A_{3000, \rm min}$. This suggests that geometry effects (e.g. the spatial distribution of dust and stars and the line of sight considered) play a major role in shaping the attenuation curve in the simulated MC. 

\subsection{Main UV and IR emission diagnostics}

\begin{figure}
    \centering
    \includegraphics[width=0.49\textwidth]{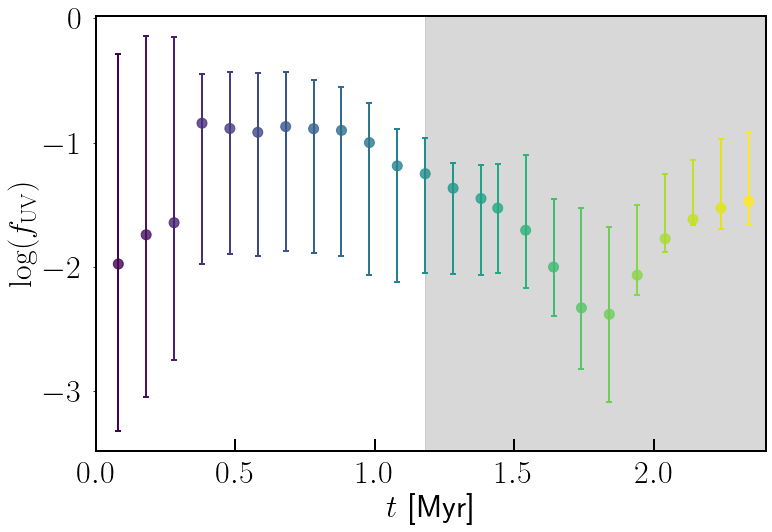}
    \hfill
    \caption{Evolution of the UV ($0.1-0.3~\mum$) escape fraction for the simulated MC. Filled circles indicate the median escape fraction among the different l.o.s, whereas errorbars bracket the minimum and maximum escape fractions. 
    The shaded area highlights the period when the cloud is dispersed according to the radiation-pressure model from \citetalias{Sommovigo:2020} (cfr. with Sec.~\ref{sec:S20_model}.) 
    \label{fig:UV_escape}
    }
\end{figure}

\begin{figure*}
    \centering
    \hfill
     \includegraphics[width=0.9\textwidth]{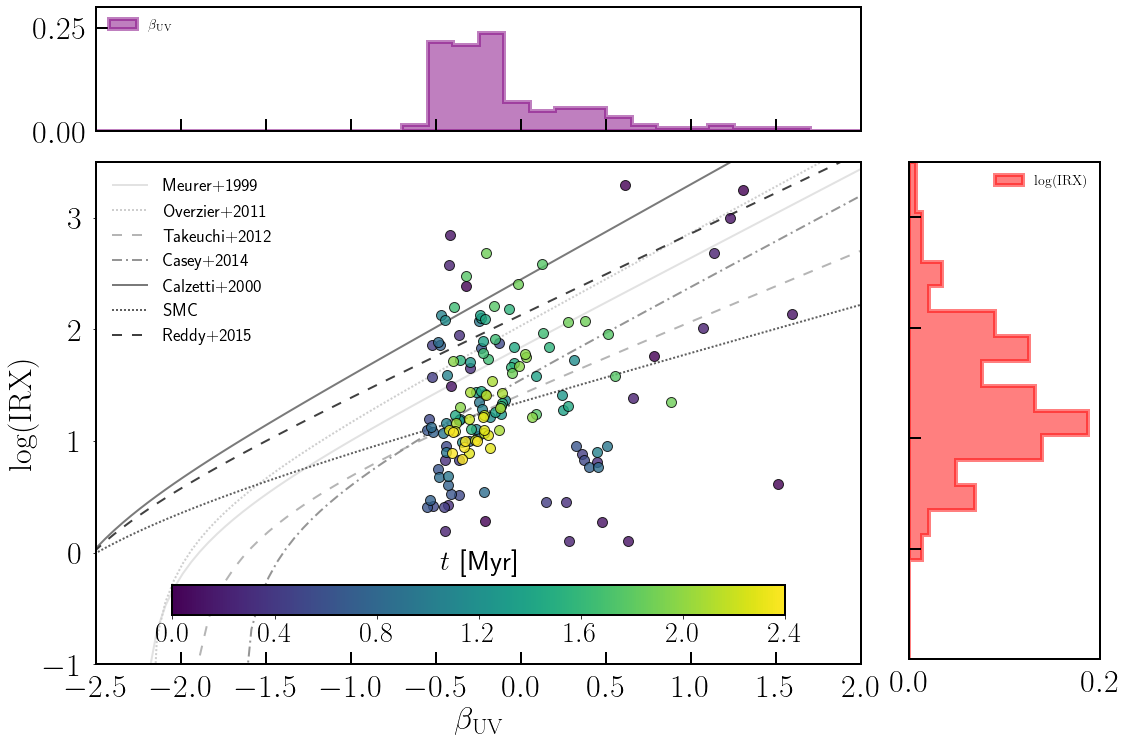}
    \caption{
    IRX-$\beta$ relation for the reference MC. Each circle corresponds to a different line of sight of a snapshot, color-coded according to the time since the onset of star formation. The upper and right insets show the 1-D PDF of $\beta_{UV}$ and IRX, respectively.
    We also show with grey lines empirical relations obtained from samples of local galaxies \citep{Meurer1999ApJ, Overzier2011ApJ, Takeuchi2012ApJ, Casey2014ApJ796}, the ones computed by \citet{Reddy2018ApJ} assuming SMC extinction, the \citet{Calzetti:2000} and \citet{Reddy2015ApJ} dust attenuation.
    \label{fig:IRXbeta_evolution}
    }
\end{figure*}

We further investigate the cloud emission properties, by studying how much radiation in the UV band effectively escape the cloud. In Fig.~\ref{fig:UV_escape} we show the evolution of the UV escape fraction $\fUV = L_{\rm UV}/L_{\rm UV, int}$. 
In the first $t \sim 0.3$~Myr after the beginning of star formation, $\fUV$ shows a variability of $3$ orders of magnitude with the l.o.s, spanning a range from $10^{-3}$ to $10^{-1}$, with a median value of $10^{-2}$. This large scatter is mainly due the variance associated with the low number of stars in place at this stage. 
At $t \sim 0.3$~Myr, the UV escape fraction increases by an order of magnitude with a median value of $\fUV \simeq 10^{-1}$. This increase happens together with the rise in the UV intrinsic luminosity (cfr. with Fig.~\ref{fig:lum_history}), suggesting that it is associated with newly born stars that experience an average lower attenuation. The higher number of stars also reduces the scatter in $\fUV$ to $1.5$~dex. 
The UV escape fraction remains constant up $t \simeq 1$~Myr, then it decreases slowly up to $t \simeq 1.5$~Myr, with also a lower scatter between different l.o.s.. Between $1.5 \lesssim t \lesssim 1.8$~Myr, it drops quickly to $3 \times 10^{-3}$ and also the variation between l.o.s. increases.
At $t \gtrsim 1.8$~Myr, the cloud begins to photo-evaporate, and this is clearly reflected in the increase of the UV escape fraction, which rises up to $3 \times 10^{-2}$ at the end of the simulation, when $\approx 1/3$ of the initial gas (and dust) mass is retained. Meanwhile, the scatter between different l.o.s. is reduced, as most of the star formation is concentrated in the center and geometry effects play a minor role (cfr. with Fig.~\ref{fig:cloud_morphology_135}).

In the context of galaxy evolution, an important diagnostic tool to quantify the conversion of UV photons into dust thermal emission in the IR band is the IRX-$\beta$ relation, which describes the infrared excess ${\rm IRX} \equiv L_{\rm IR}/L_{\rm UV}$ as a function of the UV slope $\beta_{\rm UV}$.
In Fig.~\ref{fig:IRXbeta_evolution} we show the IRX-$\beta$ relation\footnote{For the computation of the infrared excess IRX, $L_{\rm IR}$ is integrated between $8-1000~\mum$ and $L_{\rm UV}$ between $0.1-0.3~\mum$. The UV slope $\beta_{\rm UV}$ is calculated between $1600-2500~\angstrom$~ assuming $L_\lambda \propto \lambda^{\beta_{\rm UV}}$, as detailed in footnote\footref{foot_betaUV}.} for the simulated MC at different times for each of the l.o.s. considered in the RT post-process.
The MC occupies a wide range of positions in the IRX-$\beta$ diagram, with $-0.5 \lesssim \beta_{\rm UV} \lesssim 2$ and $0 \lesssim \log(\rm IRX) \lesssim 3.5$. In most cases, the MC appears with a $-0.5 < \beta_{\rm UV} < 0$ and $0.5 < \log(\rm IRX) < 2$. 
Interestingly, the position of the MC at the similar times (or even at the same age) changes dramatically depending on the l.o.s., in particular at early stages ($t \lesssim 0.5$~Myr). Instead, after the cloud starts to disperse ($t \gtrsim 1.8$~Myr) it settles at $-0.5 < \beta_{\rm UV} < 0$ and $\log(\rm IRX) = 1$.
We also show for comparison several empirically or theoretically derived relations: the fit to local starburst galaxies by \citet{Meurer1999ApJ}; the re-examinations of the previous relation with aperture-corrected data by \citet{Overzier2011ApJ} and \citet{Takeuchi2012ApJ}; a fit to local galaxies by \citep{Casey2014ApJ796} with a large dynamical range and aperture-corrected data; the relations computed by \citet{Reddy2018ApJ} assuming an SMC extinction, the Calzetti \citep{Calzetti:2000} attenuation and the \citet{Reddy2015ApJ} attenuation.

\section{Summary and conclusions} \label{sec:conclusions}

In this work, we presented Monte Carlo radiative transfer (RT) calculations with the code \code{SKIRT} \citep{CampsBaes2015, CampsBaes2020} performed to post-process the RHD simulation from \citet[][hereafter \citetalias{Decataldo2020}]{Decataldo2020}, which follows the evolution of a local Molecular Cloud (MC) with gas mass $M_{\rm gas}\approx 10^5~\msun$ down to a spatial resolution of $0.06$~pc.
We investigated the physical properties (e.g. cloud morphology, dust absorption, dust temperature) and multi-wavelength emission properties (e.g. Spectral Energy Distribution, UV escape fraction, attenuation curve), using the essential, physically-motivated, MC emission models from \citet[][hereafter \citetalias{Sommovigo:2020}]{Sommovigo:2020} as a baseline benchmark, in particularly adopting the \emph{homogeneous} dust distribution and \emph{radiation pressure} model as bracketing cases.
This work represents a starting point toward the development of sub-grid prescriptions to describe the structure and the emission of unresolved structures in cosmological hydro-dynamic simulations of galaxy formation.
We also test our results against similar works from the literature and observations of local MCs. 

Star formation rate (SFR) in the simulated MC proceeds in a bursty fashion, overall increasing from $\approx 10^{-3}~\msunyr$ to $0.15~\msunyr$ at $t\simeq 1.8$~Myr, when the cloud starts to photo-evaporate, due to stellar feedback mechanisms, such as photo-ionization, winds and radiation pressure. (Fig.~\ref{fig:sfr_history}).
In our analysis, we first focused on the cloud properties at $t\simeq 1.1$~Myr after the beginning of star formation, which we refer to as our \emph{reference} cloud. We then investigated how the physical and emission properties change throughout the cloud evolution. Our main results are the following.

In the reference MC, the stellar component ($M_\star \sim 2.3 \times 10^3~\msun$, a $\SFR \sim 0.01~\msunyr$) is more compact than the dust component ($M_{\rm dust} = 6.4~\times 10^2~\msun$), with the $90\%$ of the total dust and stellar mass enclosed in $r_{*,90}\simeq 7.4$~pc and $r_{\rm dust, 90} \simeq 13$~pc respectively. In the \citetalias{Sommovigo:2020} homogeneous model the dust mass is less concentrated (by a factor of $\approx 10$) in the inner $\sim 10$~pc, whereas in the radiation-pressure model the central regions are almost deprived of their content (the dust mass retained in the cloud at this stage in this model is $1.7~\times 10^2~\msun$). However, ($r_{\rm dust, 90}^{\rm S20 \ hom} \approx r_{\rm dust, 90}^{\rm S20 \ RP} \approx 15$~pc, comparable to the simulated MC (Fig.~\ref{fig:dust_mass_density_profile}).

The median UV optical depth computed from the dust distribution at the edge of the reference cloud is $\tau_{UV} \sim 30$, with a $0.5$~dex variation between different lines of sight. However, the presence of low-density channels and the intermixed distribution between dust and stars result in an effective UV opacity between $\tau_{\rm UV, eff} = 2.0-4.6$ (Fig.~\ref{fig:dust_tau_profile}, top panel). In the analytic model for an homogeneous cloud by \citetalias{Sommovigo:2020}, the optical depth grows as $\tau_{\rm UV}^{\rm S20 hom} \propto r$, slightly less steep than in the median profile of the simulated MC, reaching $\tau_{\rm UV}^{\rm S20 hom} \sim 10$ at the cloud edge. In radiation-pressure model, the UV optical depth remains negligible in the innermost regions due to the dust removal, and it reaches $\tau_{\rm UV}^{\rm S20 \ RP} \sim 4$ at the cloud edge. Thus we find the homogeneous model better reproducing the profile of the physical UV optical depth, whereas the RP model provides the best estimate of the effective UV optical depth at the edge of the cloud.
The reference MC is always optically thin ($\tau_{\rm IR} < 0.1$) in the IR among all the considered lines of sight, both in the simulation and analytical models (Fig.~\ref{fig:dust_tau_profile}, bottom panel). 

Dust attenuation suppresses $\gtrsim 90\%$ of the UV flux emitted by stars at this stage, with a scatter of one order of magnitude in emission depending on the line of sight (l.o.s; Fig.~\ref{fig:cloud_SED}), as a result of the complex cloud morphology, which results in most of the UV emission coming from stars whose stellar feedback cleared out the surrounding gas (Fig.~\ref{fig:cloud_morphology}). 
The total FIR emission in the analytical models is the same as in the simulated MC, because of the LTE assumption. However, the IR emission in the three models peak at different wavelengths: $\lambda_{\rm peak}=80~\mum$ in the simulation, $\lambda_{\rm peak}=40~\mum$ in the homogeneous model and $\lambda_{\rm peak}=55~\mum$ in the radiation pressure model (Fig.~\ref{fig:cloud_SED}). This difference depends on the specific dust temperature distributions in the three models. 

We compute the Probability Distribution Function (PDF) of the dust component by weighting in mass and in luminosity the contribution of each resolution element. For the reference cloud, the \emph{mass-weighted} distribution (tracing the bulk of the dust component) settles at ${\langle T_{\rm d} \rangle}_{\rm M} = 25.9^{+7.0}_{-3.8}$~K, similar to the \citetalias{Sommovigo:2020} radiation-pressure model ${\langle T_{\rm d} \rangle}_{\rm M} = 26.1^{+14.4}_{-11.9}$~K, whereas in the homogeneous model it settles to lower temperatures ${\langle T_{\rm d}\rangle}_{\rm M}$. The \emph{luminosity-weighted} distribution (tracing the warmest dusty regions) is ${\langle T_{\rm d} \rangle}_{\rm L} = 74.0^{+247}_{-38.5}$~K in the reference MC, with multiple peaks at $T>100$~K, corresponding to small pockets of dust very close to a luminous star (Fig.~\ref{fig:dust_temperature_profile}). In the \citetalias{Sommovigo:2020} homogeneous model, it is higher (${\langle T_{\rm d} \rangle}_{\rm L} = 94.2^{+50.6}_{-35.9}$~K) due to the contribution of the central regions directly irradiated by stars without experiencing any shield effect, whereas in the radiation-pressure model it has a narrow range ${\langle T_{\rm d} \rangle}_{\rm L} = 59.8^{+11.6}_{-15.9}$~K, because of the geometrical dilution due to most of the dust being pushed at large distances from the center (Fig.~\ref{fig:Tdust_PDF}).

During the cloud evolution, the mass-weighted dust temperature tends to follow the evolution of the star formation within the MC. It remains almost constant at $\approx 20-30$~K during the first $1$~Myr, grows mildly up to $\approx 52$~K at $t\simeq 1.8$~Myr, and then slowly decreases when the star formation rate in the cloud drops because of the cloud photo-ionization. The \citetalias{Sommovigo:2020} homogeneous model does not quite capture this trend, as \TdMWS20{} remains constant at $10-20$~K up to $t\simeq 1.8$~Myr and then it rises up to $\approx 55$~K. Instead in the RP model \TdMWS20{} evolves similarly to the simulated \TdMW{} up to $t\simeq 1.5$~Myr, after which the dust mass loss due to radiation pressure becomes relevant in the analytic model. \TdLW{} remains almost constant at $40-50$~K during the first $1$~Myr and then grows up to the end of the simulation, reaching $\approx 200$~K. Instead, in the homogeneous model \TdLWS20{} starts from $\approx 56$~K and moderately rises up to $\approx 150$~K at $t\simeq 1.8$~Myr, then dropping after the cloud starts to photo-evaporate, reaching $\approx 137$~K at the end of the simulation. In the radiation-pressure model \TdLWS20{} presents a very mild evolution throughout the cloud lifetime, starting at $40-50$~K and slowly growing up to $\approx 70$~K up to the cloud photo-evaporation, as no dust shells that experience a strong radiation field are present because radiation pressure pushes the dust away from the central stars (Fig.~\ref{fig:Tdust_evolution}).

The cloud emission changes significantly across its lifetime as a result of the complex evolution of its absorption properties and its star formation history. The UV ($0.1-0.3~\mum$) radiation produced by stars increases following loosely the average trend of the star formation rate (SFR) up to the cloud photo-evaporation, growing from $\sim 10^5~\lsun$ to $3.7\times 10^7~\lsun$. When the cloud photo-evaporates it remains constant despite the drop in SFR, as few massive stars are still seeded (Fig.~\ref{fig:lum_history}).

The dust-obscured UV emission from the cloud has a more complex evolution (Fig.~\ref{fig:cloud_SED_evolution}) due to the change in its absorption properties. At the early stages ($t \lesssim 0.3$~Myr), only $\fUV \approx 1\%$ of the UV emission survives dust attenuation along the median line of sight, with strong variations between different l.o.s. ($\fUV \approx 0.1-10\%$). Then, the UV transmission increases by one order of magnitude ($\fUV \approx 10\%$) and it slowly decreases down to $\fUV \approx 0.3\%$ at the cloud photo-evaporation at $t\simeq 1.7$~Myr, after which several dust-free channels are open leaving UV radiation free to escape, $\fUV \approx 3\%$ (Fig.~\ref{fig:UV_escape}). 

The peak of FIR ($8-1000~\mum$) emission increases in magnitude by $\sim 3$ orders of magnitude throughout the MC evolution, and the wavelength of the peak shifts toward shorter wavelengths, from $110~\mum$ at $t \approx 0.3$~Myr to $31~\mum$ at $t \approx 2.3$~Myr (Fig.~\ref{fig:cloud_SED_evolution}). This is due to the shift of the dust temperature distribution toward higher values as the MC evolves, because the intensity of the radiation field increases as more and more stars are seeded.

The attenuation curves of the simulated cloud show an evolution from a Milky-Way attenuation curve at early times toward a flatter, featureless relation. The $2175~\angstrom~$ bump strength also varies during the cloud lifetime (Fig.~\ref{fig:cloud_Alambda_evolution}). These variations emerge despite the intrinsic dust properties are assumed to be the same throughout the cloud evolution. This suggests that geometry effects (e.g. the spatial distribution of dust and stars and the line of sight considered) play a major role in shaping the attenuation curve in the simulated MC. 

The MC position on the IRX-$\beta$ diagram changes significantly throughout its evolution, strongly depending on the l.o.s. considered and moves significantly even after short time variation. After the cloud photo-evaporation, it settles at $-0.5 < \beta_{\rm UV} < 0$ and $\log(\rm IRX) = \simeq $ (Fig.~\ref{fig:IRXbeta_evolution}). This finding suggests caution in interpreting the IRX-$\beta$ on galaxy scale, as different regions would be expected to lie in different location on the IRX-$\beta$ plane depending on the local star formation stage.

In summary, the comparison between the simulated MC and the analytical models in terms of dust temperature distribution (e.g. \TdMW{}, \TdLW{}) and cloud emission ($f_{\rm UV}$, $L_{\rm TIR}$, SED shape) prefers the inclusion of a mechanical feedback mechanism affecting the gas/dust spatial distribution (in our specific case, the radiation pressure) in order to provide a description closer to the simulations. However, the considered radiation-pressure model is not able to capture the high-temperature regions and it under-estimates the cloud dispersal time, because all the stars are assumed to be at the center. Moreover, it does not predict an UV escape fraction, as no dust-free channels are present due to the spherical symmetry. A more sophisticated numerical setup, able to predict high-temperature regions irradiated by bright stars, and handling a more complex geometry, would provide more realistic estimates of the relevant physical quantities in describing the cloud emission on sub-grid scales, e.g. \TdMW{}, \TdLW{}, $f_{\rm UV}$, $L_{\rm TIR}$.

Finally, we also compared the results from our simulations to photo-metric data of $28$ massive star-forming HII regions in the Milky-Way from \citet{BinderPovich2018ApJ...864..136B}(Fig.~\ref{fig:cloud_SED_evolution}), finding that our SEDs present a FIR peak at shorter wavelengths despite having comparable infrared luminosities. We attribute the discrepancy to the lack of a cold dust component in our simulations due to an excessive star formation efficiency. We plan to refine in the future the adopted star formation recipe, including also the impact of magnetic fields, which we expect would reduce the gas consumption and produce MCs with physical conditions similar to the local ones.

\section*{Acknowledgements}
We acknowledge usage of the Python programming language \citep{python2,python3}, Astropy \citep{astropy}, Cython \citep{cython}, Matplotlib \citep{matplotlib}, NumPy \citep{numpy}, Numba \citep{numba}, \code{pynbody} \citep{pynbody}, and, SciPy \citep{scipy}.
We gratefully acknowledge computational resources of the Center for High Performance Computing (CHPC) at Scuola Normale Superiore, Pisa (IT). 

\section*{Data availability}
The derived data generated in this research will be shared on reasonable request to the corresponding author.

\bibliographystyle{stile/aa_url}
\bibliography{file_bibliography/ref}

\appendix

\section{Cloud morphology at early and late times}

\begin{figure*}
    \centering
    \includegraphics[width=0.9\textwidth]{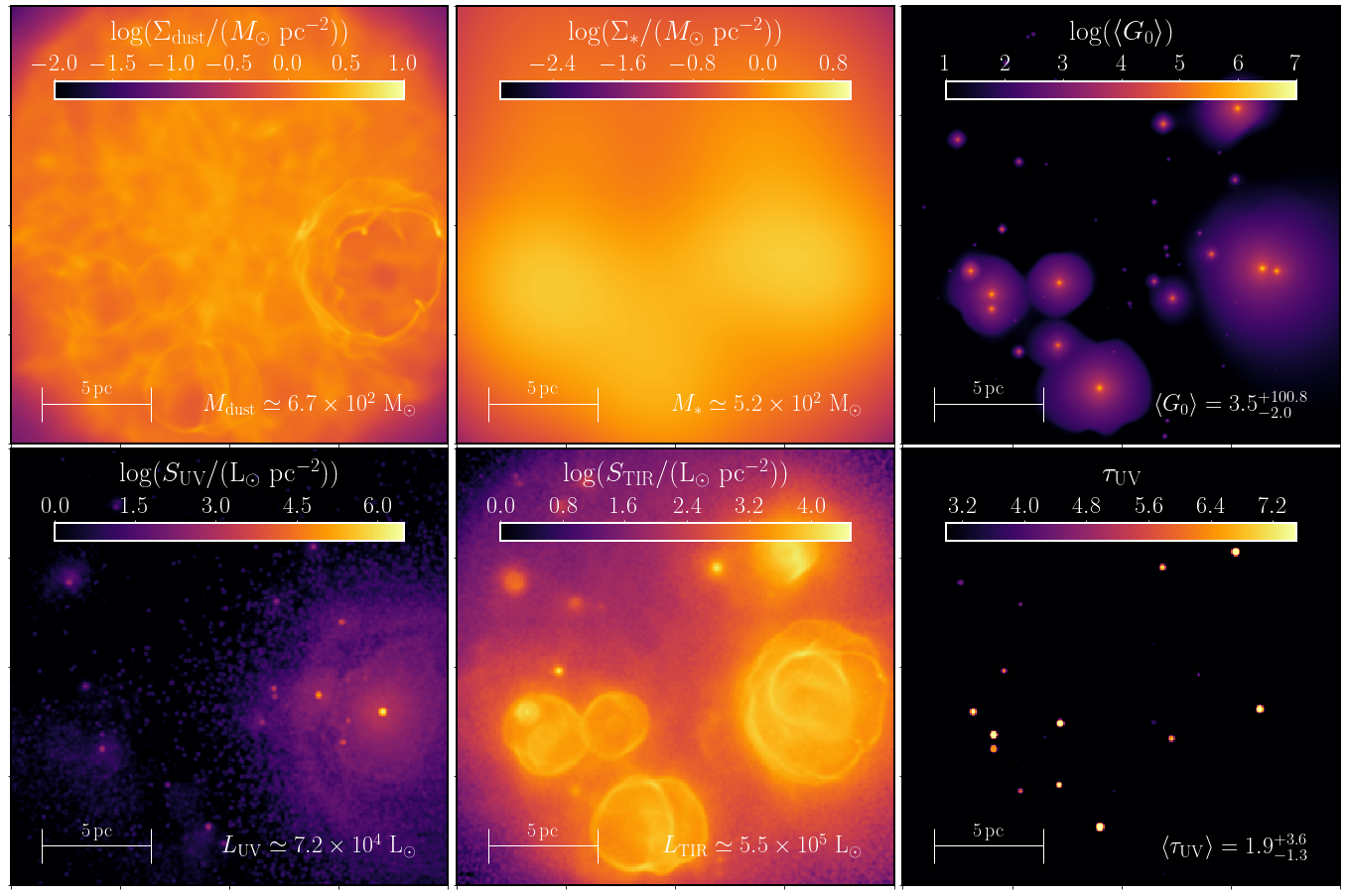}
    \caption{
    Morphology of the MC at $t\sim 0.5$~Myr (cfr. with Fig.~\ref{fig:cloud_morphology}).
    \label{fig:cloud_morphology_55}
    }
\end{figure*}

\begin{figure*}
    \centering
    \includegraphics[width=0.9\textwidth]{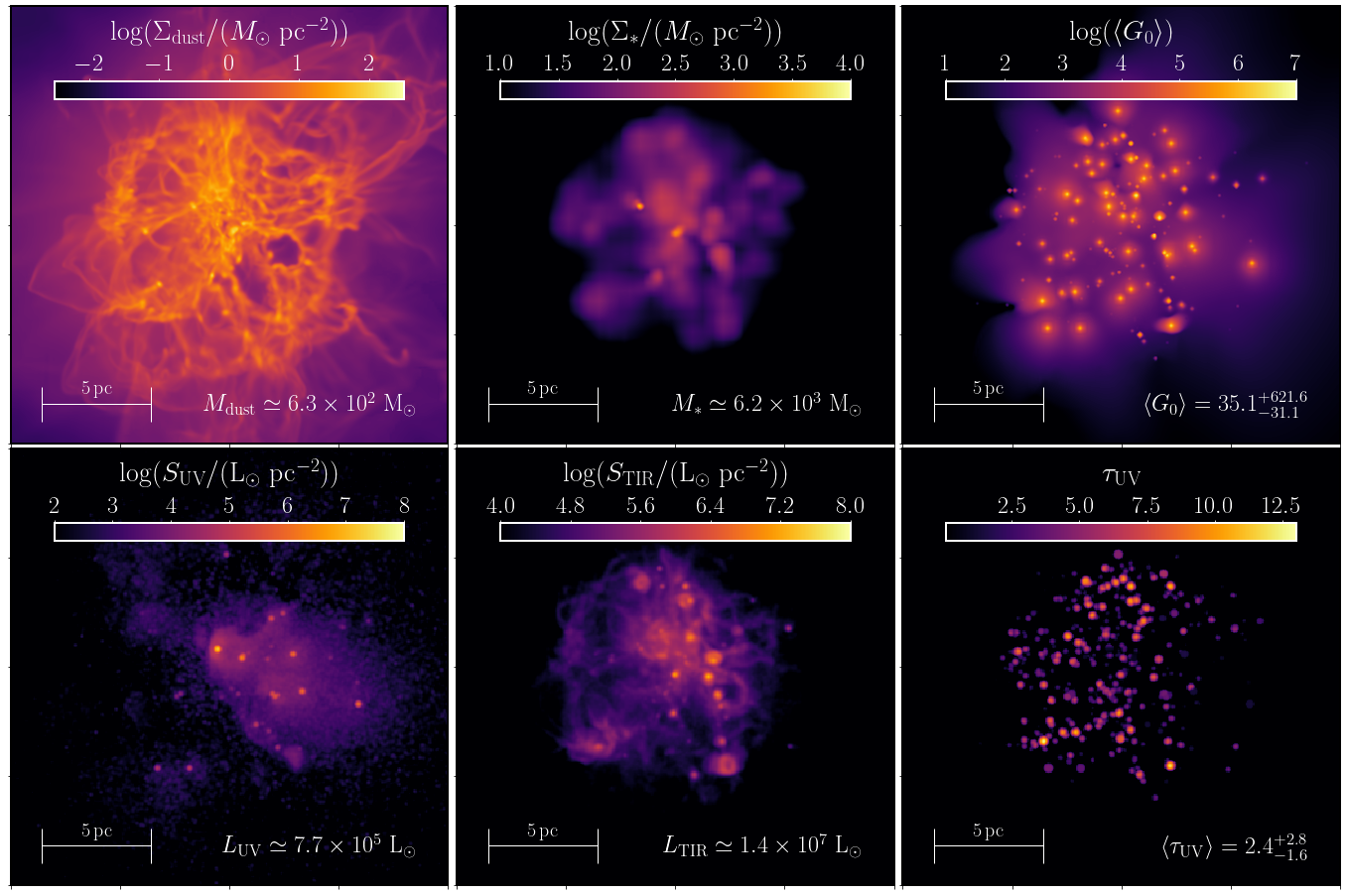}
    \caption{
    Morphology of the MC at $t\sim 1.4$~Myr (cfr. with Fig.~\ref{fig:cloud_morphology}).
    \label{fig:cloud_morphology_100}
    }
\end{figure*}

\begin{figure*}
    \centering
    \includegraphics[width=0.9\textwidth]{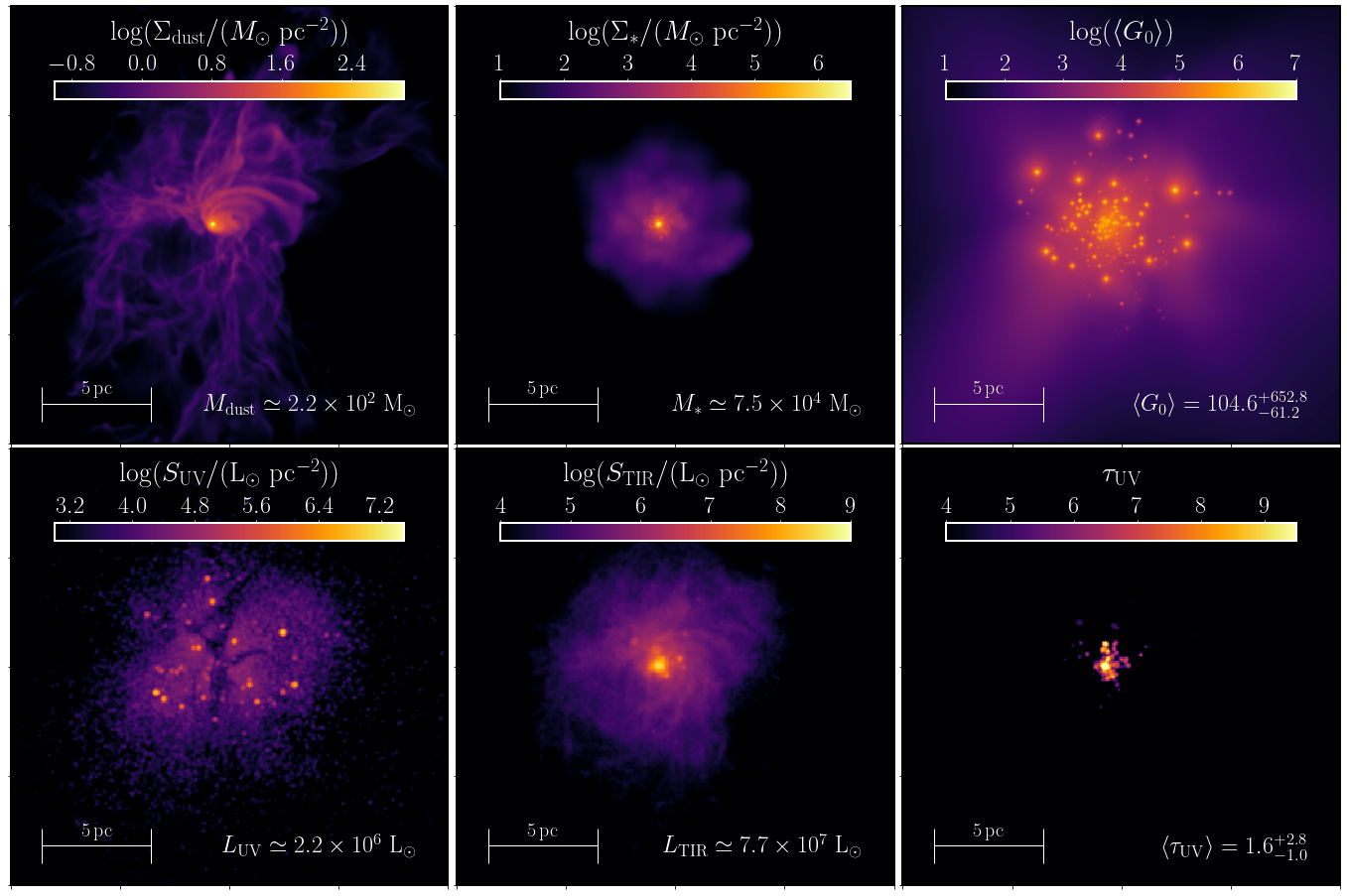}
    \caption{
    Morphology of the MC at $t\sim 2.0$~Myr (cfr. with Fig.~\ref{fig:cloud_morphology}).
    \label{fig:cloud_morphology_135}
    }
\end{figure*}

In Sec.~\ref{sec:cloud_morph} we discuss the morphology of the MC after $t \simeq 1.2$~Myr since the beginning of star formation.
In order to qualitatively illustrate the sensitivity of the the cloud morphology due to the evolutionary stage, we report the same physical quantities as in Fig.~\ref{fig:cloud_morphology} at three different epochs: i) $t\sim 0.5$~Myr (Fig.~\ref{fig:cloud_morphology_55}), when $\sim 0.7\%$ of the total final stellar mass is present, ii) $t\sim 1.4$~Myr (Fig.~\ref{fig:cloud_morphology_100}) when $\sim 3.1\%$ of the total final stellar mass is present, and iii) $t \sim 2$~Myr (Fig.~\ref{fig:cloud_morphology_135}), when the cloud is toward the end of its life and $36\%$ of the gas mass remains. 

\end{document}

%% file: include_tex/pacchetti.tex
\usepackage{txfonts}

\usepackage[english]{babel}
\usepackage{amsmath,amssymb}
\usepackage{graphicx, subfig}
\usepackage[normalem]{ulem}

\usepackage{verbatim}

\usepackage{multirow}
\usepackage{mathtools}
\usepackage{epstopdf}

\usepackage{url}
\usepackage{xcolor,color}
\usepackage{orcidlink}

\usepackage{natbib,twoopt}
\usepackage[hyphenbreaks]{breakurl}

%% file: include_tex/aa_url_patch.tex
\bibpunct{(}{)}{;}{a}{}{,}             
\definecolor{cobalt}{rgb}{0.06, 0.2, 0.65}
\hypersetup{
  colorlinks,
  citecolor=cobalt,
  linkcolor=[rgb]{0.8, 0.2, 1.0},
  urlcolor=cobalt,
}
\makeatletter
  \newcommandtwoopt{\citeads}[3][][]{\href{http://adsabs.harvard.edu/abs/#3}%
    {\def\hyper@linkstart##1##2{}%
     \let\hyper@linkend\@empty\citealp[#1][#2]{#3}}}
  \newcommandtwoopt{\citepads}[3][][]{\href{http://adsabs.harvard.edu/abs/#3}%
    {\def\hyper@linkstart##1##2{}%
     \let\hyper@linkend\@empty\citep[#1][#2]{#3}}}
  \newcommandtwoopt{\citetads}[3][][]{\href{http://adsabs.harvard.edu/abs/#3}%
    {\def\hyper@linkstart##1##2{}%
     \let\hyper@linkend\@empty\citet[#1][#2]{#3}}}
  \newcommandtwoopt{\citeyearads}[3][][]%
    {\href{http://adsabs.harvard.edu/abs/#3}
    {\def\hyper@linkstart##1##2{}%
     \let\hyper@linkend\@empty\citeyear[#1][#2]{#3}}}
\makeatother

%% file: include_tex/journals.tex
\def\apjl{ApJL}
\def\apjs{ApJS}
\def\aap{A\&A}


%% file: include_tex/definizioni.tex
\def\be{\begin{equation}} 
\def\ee{\end{equation}} 
\def\ba{\begin{eqnarray}} 
\def\ea{\end{eqnarray}}

\def\cc{\,{\rm {cm^{-3}}}} 
\def\msun{{\Msun}}

\def\HII{\hbox{H~$\scriptstyle\rm II\ $}}

  \def\cm{${\rm cm}$}

\def\gsim{\lower.5ex\hbox{\gtsima}} 
\def\lsim{\lower.5ex\hbox{\ltsima}} \def\gtsima{$\; \buildrel > \over 
\sim \;$} \def\ltsima{$\; \buildrel < \over \sim \;$} \def\prosima{$\; 
\buildrel \propto \over \sim \;$} \def\gsim{\lower.5ex\hbox{\gtsima}} 
\def\lsim{\lower.5ex\hbox{\ltsima}} 
\def\simgt{\lower.5ex\hbox{\gtsima}} 
\def\simlt{\lower.5ex\hbox{\ltsima}} 
\def\simpr{\lower.5ex\hbox{\prosima}} \def\la{\lsim} \def\ga{\gsim} 
  
 \def\gtsima{$\; \buildrel > \over \sim \;$} 
\def\ltsima{$\; \buildrel < \over \sim \;$} 
\def\gsim{\lower.5ex\hbox{\gtsima}} 
\def\lsim{\lower.5ex\hbox{\ltsima}} 
\def\simgt{\lower.5ex\hbox{\gtsima}} 
\def\simlt{\lower.5ex\hbox{\ltsima}} 
\def\simpr{\lower.5ex\hbox{\prosima}} 
\def\la{\lsim} 
\def\ga{\gsim}

\def\msun{\,{\rm \Msun}}

\def\E3{{\cal E}_{\rm g}^{III}}

\def\r12{r_{1/2}} 
\def\x12{x_{1/2}} 
\def\v12{v_{1/2}}

\def\cm2g{ {\rm cm^2} \, {\rm {g^{-1}}}}

%% file: include_tex/additional_def.tex
\def\HII{\hbox{H~$\scriptstyle\rm II $~}}

\newcommand\code[1]{\textsc{\MakeLowercase{#1}}}

\def\nh2{n_{\rm H2}}
\def\fh2{f_{\rm H2}}

\def\angstrom{\textrm{A\kern -1.3ex\raisebox{0.6ex}{$^\circ$}}}

\def\mum{\mu{\rm m}}
\def\msun{{\rm M}_{\odot}}
\def\zsun{{\rm Z}_{\odot}}
\def\lsun{{\rm L}_{\odot}}

\def\cc{{\rm cm}^{-3}}

\def\msunyr{\msun\,{\rm yr}^{-1}}

\def\surfdpc{\msun\,{\rm pc}^{-2}}
\def\surfl{\lsun\,{\rm kpc}^{-2}}

\def\SFR{{\rm SFR}}

\def\fUV{f_{\rm UV}}

\def\TdMW{${\langle T_{\rm d} \rangle}_{\rm M}$}
\def\TdLW{${\langle T_{\rm d} \rangle}_{\rm L}$}
\def\TdMWS20{${\langle T_{\rm d} \rangle}_{\rm M}^{\rm S20}$}
\def\TdLWS20{${\langle T_{\rm d} \rangle}_{\rm L}^{\rm S20}$}

%% file: include_tex/colors.tex
\makeatletter
\def\@hex@@Hex#1%
 {\if a#1A\else \if b#1B\else \if c#1C\else \if d#1D\else
  \if e#1E\else \if f#1F\else #1\fi\fi\fi\fi\fi\fi \@hex@Hex}
\makeatother

\definecolor{apcolor}{HTML}{b3003b}
\definecolor{vdcolor}{HTML}{ff0f00}
\definecolor{afcolor}{HTML}{b3443c}
\definecolor{sgcolor}{HTML}{008000}
\definecolor{sccolor}{HTML}{ffbe33}
\definecolor{fvcolor}{HTML}{ce33ff}
\definecolor{pbcolor}{HTML}{333300}
\definecolor{fdcolor}{HTML}{ff6600}
\definecolor{ddcolor}{HTML}{26adf0}
\definecolor{todo}{HTML}{bc4be9}

%% file: main.bbl
\begin{thebibliography}{107}
\expandafter\ifx\csname natexlab\endcsname\relax\def\natexlab#1{#1}\fi

\bibitem[{{Astropy Collaboration} {et~al.}(2013){Astropy Collaboration},
  {Robitaille}, {Tollerud}, {Greenfield}, {Droettboom}, {Bray}, {Aldcroft},
  {Davis}, {Ginsburg}, {Price-Whelan}, {Kerzendorf}, {Conley}, {Crighton},
  {Barbary}, {Muna}, {Ferguson}, {Grollier}, {Parikh}, {Nair}, {Unther},
  {Deil}, {Woillez}, {Conseil}, {Kramer}, {Turner}, {Singer}, {Fox}, {Weaver},
  {Zabalza}, {Edwards}, {Azalee Bostroem}, {Burke}, {Casey}, {Crawford},
  {Dencheva}, {Ely}, {Jenness}, {Labrie}, {Lim}, {Pierfederici}, {Pontzen},
  {Ptak}, {Refsdal}, {Servillat}, \& {Streicher}}]{astropy}
{Astropy Collaboration}, {Robitaille}, T.~P., {Tollerud}, E.~J., {et~al.} 2013,
  \href{http://dx.doi.org/10.1051/0004-6361/201322068}{\color{magenta}\aap},
  \href{http://adsabs.harvard.edu/abs/2013A%26A...558A..33A}{558, A33}

\bibitem[{{Aubert} \& {Teyssier}(2008)}]{aubert:2008}
{Aubert}, D. \& {Teyssier}, R. 2008,
  \href{http://dx.doi.org/10.1111/j.1365-2966.2008.13223.x}{\color{magenta}\mnras},
  \href{https://ui.adsabs.harvard.edu/abs/2008MNRAS.387..295A}{387, 295}

\bibitem[{Behnel {et~al.}(2011)Behnel, Bradshaw, Citro, Dalcin, Seljebotn, \&
  Smith}]{cython}
Behnel, S., Bradshaw, R., Citro, C., {et~al.} 2011,
  \href{http://dx.doi.org/10.1109/MCSE.2010.118}{\color{magenta}Computing in
  Science Engineering}, 13, 31

\bibitem[{{Behrens} {et~al.}(2018){Behrens}, {Pallottini}, {Ferrara},
  {Gallerani}, \& {Vallini}}]{Behrens:2018}
{Behrens}, C., {Pallottini}, A., {Ferrara}, A., {Gallerani}, S., \& {Vallini},
  L. 2018,
  \href{http://dx.doi.org/10.1093/mnras/sty552}{\color{magenta}\mnras},
  \href{https://ui.adsabs.harvard.edu/abs/2018MNRAS.477..552B}{477, 552}

\bibitem[{{Binder} \& {Povich}(2018)}]{BinderPovich2018ApJ...864..136B}
{Binder}, B.~A. \& {Povich}, M.~S. 2018,
  \href{http://dx.doi.org/10.3847/1538-4357/aad7b2}{\color{magenta}\apj},
  \href{https://ui.adsabs.harvard.edu/abs/2018ApJ...864..136B}{864, 136}

\bibitem[{{Bisbas} {et~al.}(2011){Bisbas}, {W{\"u}nsch}, {Whitworth}, {Hubber},
  \& {Walch}}]{Bisbas2011}
{Bisbas}, T.~G., {W{\"u}nsch}, R., {Whitworth}, A.~P., {Hubber}, D.~A., \&
  {Walch}, S. 2011,
  \href{http://dx.doi.org/10.1088/0004-637X/736/2/142}{\color{magenta}\apj},
  \href{https://ui.adsabs.harvard.edu/abs/2011ApJ...736..142B}{736, 142}

\bibitem[{{Bovino} {et~al.}(2016){Bovino}, {Grassi}, {Capelo}, {Schleicher}, \&
  {Banerjee}}]{bovino:2016}
{Bovino}, S., {Grassi}, T., {Capelo}, P.~R., {Schleicher}, D.~R.~G., \&
  {Banerjee}, R. 2016,
  \href{http://dx.doi.org/10.1051/0004-6361/201628158}{\color{magenta}\aap},
  \href{https://ui.adsabs.harvard.edu/abs/2016A&A...590A..15B}{590, A15}

\bibitem[{{Bressan} {et~al.}(2012){Bressan}, {Marigo}, {Girardi}, {Salasnich},
  {Dal Cero}, {Rubele}, \& {Nanni}}]{bressan:2012}
{Bressan}, A., {Marigo}, P., {Girardi}, L., {et~al.} 2012,
  \href{http://dx.doi.org/10.1111/j.1365-2966.2012.21948.x}{\color{magenta}\mnras},
  \href{https://ui.adsabs.harvard.edu/abs/2012MNRAS.427..127B}{427, 127}

\bibitem[{{Brunt} {et~al.}(2009){Brunt}, {Heyer}, \& {Mac Low}}]{Brunt2009}
{Brunt}, C.~M., {Heyer}, M.~H., \& {Mac Low}, M.~M. 2009,
  \href{http://dx.doi.org/10.1051/0004-6361/200911797}{\color{magenta}\aap},
  \href{https://ui.adsabs.harvard.edu/abs/2009A&A...504..883B}{504, 883}

\bibitem[{{Butler} {et~al.}(2015){Butler}, {Tan}, \& {Van
  Loo}}]{butler:2015ApJ}
{Butler}, M.~J., {Tan}, J.~C., \& {Van Loo}, S. 2015,
  \href{http://dx.doi.org/10.1088/0004-637X/805/1/1}{\color{magenta}\apj},
  \href{https://ui.adsabs.harvard.edu/abs/2015ApJ...805....1B}{805, 1}

\bibitem[{{Calzetti} {et~al.}(2000){Calzetti}, {Armus}, {Bohlin}, {Kinney},
  {Koornneef}, \& {Storchi-Bergmann}}]{Calzetti:2000}
{Calzetti}, D., {Armus}, L., {Bohlin}, R.~C., {et~al.} 2000,
  \href{http://dx.doi.org/10.1086/308692}{\color{magenta}\apj},
  \href{https://ui.adsabs.harvard.edu/abs/2000ApJ...533..682C}{533, 682}

\bibitem[{{Camps} \& {Baes}(2015)}]{CampsBaes2015}
{Camps}, P. \& {Baes}, M. 2015,
  \href{http://dx.doi.org/10.1016/j.ascom.2014.10.004}{\color{magenta}Astronomy
  and Computing},
  \href{https://ui.adsabs.harvard.edu/abs/2015A&C.....9...20C}{9, 20}

\bibitem[{{Camps} \& {Baes}(2020)}]{CampsBaes2020}
{Camps}, P. \& {Baes}, M. 2020,
  \href{http://dx.doi.org/10.1016/j.ascom.2020.100381}{\color{magenta}Astronomy
  and Computing},
  \href{https://ui.adsabs.harvard.edu/abs/2020A&C....3100381C}{31, 100381}

\bibitem[{{Camps} {et~al.}(2016){Camps}, {Trayford}, {Baes}, {Theuns},
  {Schaller}, \& {Schaye}}]{Camps2016}
{Camps}, P., {Trayford}, J.~W., {Baes}, M., {et~al.} 2016,
  \href{http://dx.doi.org/10.1093/mnras/stw1735}{\color{magenta}\mnras},
  \href{https://ui.adsabs.harvard.edu/abs/2016MNRAS.462.1057C}{462, 1057}

\bibitem[{{Casey} {et~al.}(2014){Casey}, {Scoville}, {Sanders}, {Lee},
  {Cooray}, {Finkelstein}, {Capak}, {Conley}, {De Zotti}, {Farrah}, {Fu}, {Le
  Floc'h}, {Ilbert}, {Ivison}, \& {Takeuchi}}]{Casey2014ApJ796}
{Casey}, C.~M., {Scoville}, N.~Z., {Sanders}, D.~B., {et~al.} 2014,
  \href{http://dx.doi.org/10.1088/0004-637X/796/2/95}{\color{magenta}\apj},
  \href{https://ui.adsabs.harvard.edu/abs/2014ApJ...796...95C}{796, 95}

\bibitem[{{Castelli} \& {Kurucz}(2003)}]{castelli:2003}
{Castelli}, F. \& {Kurucz}, R.~L. 2003, in Modelling of Stellar Atmospheres,
  ed. N.~{Piskunov}, W.~W. {Weiss}, \& D.~F. {Gray}, Vol. 210,
  \href{https://ui.adsabs.harvard.edu/abs/2003IAUS..210P.A20C}{A20}

\bibitem[{{Castor} {et~al.}(1975){Castor}, {Abbott}, \&
  {Klein}}]{Castor1975ApJ}
{Castor}, J.~I., {Abbott}, D.~C., \& {Klein}, R.~I. 1975,
  \href{http://dx.doi.org/10.1086/153315}{\color{magenta}\apj},
  \href{https://ui.adsabs.harvard.edu/abs/1975ApJ...195..157C}{195, 157}

\bibitem[{Ceverino {et~al.}(2017)Ceverino, Glover, \&
  Klessen}]{Ceverino_Glover_Klessen_2017}
Ceverino, D., Glover, S. C.~O., \& Klessen, R.~S. 2017,
  \href{http://dx.doi.org/10.1093/mnras/stx1386}{\color{magenta}Monthly Notices
  of the Royal Astronomical Society}, 470, 2791–2798

\bibitem[{{Chevance} {et~al.}(2020){Chevance}, {Kruijssen}, {Hygate},
  {Schruba}, {Longmore}, {Groves}, {Henshaw}, {Herrera}, {Hughes}, {Jeffreson},
  {Lang}, {Leroy}, {Meidt}, {Pety}, {Razza}, {Rosolowsky}, {Schinnerer},
  {Bigiel}, {Blanc}, {Emsellem}, {Faesi}, {Glover}, {Haydon}, {Ho}, {Kreckel},
  {Lee}, {Liu}, {Querejeta}, {Saito}, {Sun}, {Usero}, \&
  {Utomo}}]{Chevance2020MNRAS.493.2872C}
{Chevance}, M., {Kruijssen}, J.~M.~D., {Hygate}, A. P.~S., {et~al.} 2020,
  \href{http://dx.doi.org/10.1093/mnras/stz3525}{\color{magenta}\mnras},
  \href{https://ui.adsabs.harvard.edu/abs/2020MNRAS.493.2872C}{493, 2872}

\bibitem[{{Churchwell} {et~al.}(2009){Churchwell}, {Babler}, {Meade},
  {Whitney}, {Benjamin}, {Indebetouw}, {Cyganowski}, {Robitaille}, {Povich},
  {Watson}, \& {Bracker}}]{Churchwell2009PASP}
{Churchwell}, E., {Babler}, B.~L., {Meade}, M.~R., {et~al.} 2009,
  \href{http://dx.doi.org/10.1086/597811}{\color{magenta}\pasp},
  \href{https://ui.adsabs.harvard.edu/abs/2009PASP..121..213C}{121, 213}

\bibitem[{{Cochrane} {et~al.}(2019){Cochrane}, {Hayward},
  {Angl{\'e}s-Alc{\'a}zar}, {Lotz}, {Parsotan}, {Ma}, {Keres}, {Feldmann},
  {Faucher-Gigu{\`e}re}, \& {Hopkins}}]{Cochrane2019MNRAS}
{Cochrane}, R.~K., {Hayward}, C.~C., {Angl{\'e}s-Alc{\'a}zar}, D., {et~al.}
  2019, \href{http://dx.doi.org/10.1093/mnras/stz1736}{\color{magenta}\mnras},
  \href{https://ui.adsabs.harvard.edu/abs/2019MNRAS.488.1779C}{488, 1779}

\bibitem[{{Colombo} {et~al.}(2014){Colombo}, {Hughes}, {Schinnerer}, {Meidt},
  {Leroy}, {Pety}, {Dobbs}, {Garc{\'\i}a-Burillo}, {Dumas}, {Thompson},
  {Schuster}, \& {Kramer}}]{Colombo2014a}
{Colombo}, D., {Hughes}, A., {Schinnerer}, E., {et~al.} 2014,
  \href{http://dx.doi.org/10.1088/0004-637X/784/1/3}{\color{magenta}\apj},
  \href{https://ui.adsabs.harvard.edu/abs/2014ApJ...784....3C}{784, 3}

\bibitem[{{Dale} {et~al.}(2005){Dale}, {Bonnell}, {Clarke}, \&
  {Bate}}]{Dale2005MNRAS.358..291D}
{Dale}, J.~E., {Bonnell}, I.~A., {Clarke}, C.~J., \& {Bate}, M.~R. 2005,
  \href{http://dx.doi.org/10.1111/j.1365-2966.2005.08806.x}{\color{magenta}\mnras},
  \href{https://ui.adsabs.harvard.edu/abs/2005MNRAS.358..291D}{358, 291}

\bibitem[{{Decataldo} {et~al.}(2020){Decataldo}, {Lupi}, {Ferrara},
  {Pallottini}, \& {Fumagalli}}]{Decataldo2020}
{Decataldo}, D., {Lupi}, A., {Ferrara}, A., {Pallottini}, A., \& {Fumagalli},
  M. 2020,
  \href{http://dx.doi.org/10.1093/mnras/staa2326}{\color{magenta}\mnras},
  \href{https://ui.adsabs.harvard.edu/abs/2020MNRAS.497.4718D}{497, 4718}

\bibitem[{{Decataldo} {et~al.}(2019){Decataldo}, {Pallottini}, {Ferrara},
  {Vallini}, \& {Gallerani}}]{decataldo:2019}
{Decataldo}, D., {Pallottini}, A., {Ferrara}, A., {Vallini}, L., \&
  {Gallerani}, S. 2019,
  \href{http://dx.doi.org/10.1093/mnras/stz1527}{\color{magenta}\mnras},
  \href{https://ui.adsabs.harvard.edu/abs/2019MNRAS.487.3377D}{487, 3377}

\bibitem[{{Deparis} {et~al.}(2019){Deparis}, {Aubert}, {Ocvirk}, {Chardin}, \&
  {Lewis}}]{deparis:2019}
{Deparis}, N., {Aubert}, D., {Ocvirk}, P., {Chardin}, J., \& {Lewis}, J. 2019,
  \href{http://dx.doi.org/10.1051/0004-6361/201832889}{\color{magenta}\aap},
  \href{https://ui.adsabs.harvard.edu/abs/2019A&A...622A.142D}{622, A142}

\bibitem[{{Di Mascia} {et~al.}(2023){Di Mascia}, {Carniani}, {Gallerani},
  {Vito}, {Pallottini}, {Ferrara}, \& {Valentini}}]{dimascia:2023}
{Di Mascia}, F., {Carniani}, S., {Gallerani}, S., {et~al.} 2023,
  \href{http://dx.doi.org/10.1093/mnras/stac3306}{\color{magenta}\mnras},
  \href{https://ui.adsabs.harvard.edu/abs/2023MNRAS.518.3667D}{518, 3667}

\bibitem[{{Di Mascia} {et~al.}(2021){Di Mascia}, {Gallerani}, {Behrens},
  {Pallottini}, {Carniani}, {Ferrara}, {Barai}, {Vito}, \&
  {Zana}}]{DiMascia2021infrared}
{Di Mascia}, F., {Gallerani}, S., {Behrens}, C., {et~al.} 2021,
  \href{http://dx.doi.org/10.1093/mnras/stab528}{\color{magenta}\mnras},
  \href{https://ui.adsabs.harvard.edu/abs/2021MNRAS.503.2349D}{503, 2349}

\bibitem[{{Dobbs}(2015)}]{Dobbs2015MNRAS.447}
{Dobbs}, C.~L. 2015,
  \href{http://dx.doi.org/10.1093/mnras/stu2585}{\color{magenta}\mnras},
  \href{https://ui.adsabs.harvard.edu/abs/2015MNRAS.447.3390D}{447, 3390}

\bibitem[{{Draine}(2011)}]{draine:2011}
{Draine}, B.~T. 2011,
  \href{http://dx.doi.org/10.1088/0004-637X/732/2/100}{\color{magenta}\apj},
  \href{https://ui.adsabs.harvard.edu/abs/2011ApJ...732..100D}{732, 100}

\bibitem[{{Evans} {et~al.}(2009){Evans}, {Dunham}, {J{\o}rgensen}, {Enoch},
  {Mer{\'\i}n}, {van Dishoeck}, {Alcal{\'a}}, {Myers}, {Stapelfeldt}, {Huard},
  {Allen}, {Harvey}, {van Kempen}, {Blake}, {Koerner}, {Mundy}, {Padgett}, \&
  {Sargent}}]{Evans2009ApJS..181..321E}
{Evans}, Neal~J., I., {Dunham}, M.~M., {J{\o}rgensen}, J.~K., {et~al.} 2009,
  \href{http://dx.doi.org/10.1088/0067-0049/181/2/321}{\color{magenta}\apjs},
  \href{https://ui.adsabs.harvard.edu/abs/2009ApJS..181..321E}{181, 321}

\bibitem[{{Federrath} \& {Klessen}(2012)}]{federrath:2012}
{Federrath}, C. \& {Klessen}, R.~S. 2012,
  \href{http://dx.doi.org/10.1088/0004-637X/761/2/156}{\color{magenta}\apj},
  \href{https://ui.adsabs.harvard.edu/abs/2012ApJ...761..156F}{761, 156}

\bibitem[{{Feng} {et~al.}(2016){Feng}, {Di-Matteo}, {Croft}, {Bird},
  {Battaglia}, \& {Wilkins}}]{Feng2016MNRAS}
{Feng}, Y., {Di-Matteo}, T., {Croft}, R.~A., {et~al.} 2016,
  \href{http://dx.doi.org/10.1093/mnras/stv2484}{\color{magenta}\mnras},
  \href{https://ui.adsabs.harvard.edu/abs/2016MNRAS.455.2778F}{455, 2778}

\bibitem[{{Forbrich} {et~al.}(2009){Forbrich}, {Lada}, {Muench}, {Alves}, \&
  {Lombardi}}]{Forbrich2009ApJ...704..292F}
{Forbrich}, J., {Lada}, C.~J., {Muench}, A.~A., {Alves}, J., \& {Lombardi}, M.
  2009,
  \href{http://dx.doi.org/10.1088/0004-637X/704/1/292}{\color{magenta}\apj},
  \href{https://ui.adsabs.harvard.edu/abs/2009ApJ...704..292F}{704, 292}

\bibitem[{{Furlanetto} \& {Mirocha}(2022)}]{furlanetto:2022}
{Furlanetto}, S.~R. \& {Mirocha}, J. 2022,
  \href{http://dx.doi.org/10.1093/mnras/stac310}{\color{magenta}\mnras},
  \href{https://ui.adsabs.harvard.edu/abs/2022MNRAS.511.3895F}{511, 3895}

\bibitem[{{Garc{\'\i}a} {et~al.}(2014){Garc{\'\i}a}, {Bronfman}, {Nyman},
  {Dame}, \& {Luna}}]{Garcia2014ApJS..212....2G}
{Garc{\'\i}a}, P., {Bronfman}, L., {Nyman}, L.-A., {Dame}, T.~M., \& {Luna}, A.
  2014,
  \href{http://dx.doi.org/10.1088/0067-0049/212/1/2}{\color{magenta}\apjs},
  \href{https://ui.adsabs.harvard.edu/abs/2014ApJS..212....2G}{212, 2}

\bibitem[{{Gatto} {et~al.}(2017){Gatto}, {Walch}, {Naab}, {Girichidis},
  {W{\"u}nsch}, {Glover}, {Klessen}, {Clark}, {Peters}, {Derigs}, {Baczynski},
  \& {Puls}}]{gatto:2017}
{Gatto}, A., {Walch}, S., {Naab}, T., {et~al.} 2017,
  \href{http://dx.doi.org/10.1093/mnras/stw3209}{\color{magenta}\mnras},
  \href{https://ui.adsabs.harvard.edu/abs/2017MNRAS.466.1903G}{466, 1903}

\bibitem[{{Geen} {et~al.}(2021){Geen}, {Bieri}, {Rosdahl}, \& {de
  Koter}}]{geen:2021}
{Geen}, S., {Bieri}, R., {Rosdahl}, J., \& {de Koter}, A. 2021,
  \href{http://dx.doi.org/10.1093/mnras/staa3705}{\color{magenta}\mnras},
  \href{https://ui.adsabs.harvard.edu/abs/2021MNRAS.501.1352G}{501, 1352}

\bibitem[{{Grassi} {et~al.}(2014){Grassi}, {Bovino}, {Schleicher}, {Prieto},
  {Seifried}, {Simoncini}, \& {Gianturco}}]{grassi:2014}
{Grassi}, T., {Bovino}, S., {Schleicher}, D.~R.~G., {et~al.} 2014,
  \href{http://dx.doi.org/10.1093/mnras/stu114}{\color{magenta}\mnras},
  \href{https://ui.adsabs.harvard.edu/abs/2014MNRAS.439.2386G}{439, 2386}

\bibitem[{{Guillet} \& {Teyssier}(2011)}]{guillet:2011}
{Guillet}, T. \& {Teyssier}, R. 2011,
  \href{http://dx.doi.org/10.1016/j.jcp.2011.02.044}{\color{magenta}Journal of
  Computational Physics},
  \href{https://ui.adsabs.harvard.edu/abs/2011JCoPh.230.4756G}{230, 4756}

\bibitem[{{Haid} {et~al.}(2018){Haid}, {Walch}, {Seifried}, {W{\"u}nsch},
  {Dinnbier}, \& {Naab}}]{haid:2018}
{Haid}, S., {Walch}, S., {Seifried}, D., {et~al.} 2018,
  \href{http://dx.doi.org/10.1093/mnras/sty1315}{\color{magenta}\mnras},
  \href{https://ui.adsabs.harvard.edu/abs/2018MNRAS.478.4799H}{478, 4799}

\bibitem[{{Haworth} {et~al.}(2018){Haworth}, {Glover}, {Koepferl}, {Bisbas}, \&
  {Dale}}]{Haworth2018NewAR..82....1H}
{Haworth}, T.~J., {Glover}, S. C.~O., {Koepferl}, C.~M., {Bisbas}, T.~G., \&
  {Dale}, J.~E. 2018,
  \href{http://dx.doi.org/10.1016/j.newar.2018.06.001}{\color{magenta}\nar},
  \href{https://ui.adsabs.harvard.edu/abs/2018NewAR..82....1H}{82, 1}

\bibitem[{{Hennebelle} {et~al.}(2022){Hennebelle}, {Lebreuilly}, {Colman},
  {Elia}, {Fuller}, {Leurini}, {Nony}, {Schisano}, {Soler}, {Traficante},
  {Klessen}, {Molinari}, \& {Testi}}]{Hennebelle2022}
{Hennebelle}, P., {Lebreuilly}, U., {Colman}, T., {et~al.} 2022,
  \href{http://dx.doi.org/10.1051/0004-6361/202243803}{\color{magenta}\aap},
  \href{https://ui.adsabs.harvard.edu/abs/2022A&A...668A.147H}{668, A147}

\bibitem[{{Heyer} {et~al.}(2009){Heyer}, {Krawczyk}, {Duval}, \&
  {Jackson}}]{heyer:2009}
{Heyer}, M., {Krawczyk}, C., {Duval}, J., \& {Jackson}, J.~M. 2009,
  \href{http://dx.doi.org/10.1088/0004-637X/699/2/1092}{\color{magenta}\apj},
  \href{https://ui.adsabs.harvard.edu/abs/2009ApJ...699.1092H}{699, 1092}

\bibitem[{{Hopkins} {et~al.}(2018){Hopkins}, {Wetzel}, {Keres},
  {Faucher-Gigu{\`e}re}, {Quataert}, {Boylan-Kolchin}, {Murray}, {Hayward},
  {Garrison-Kimmel}, {Hummels}, {Feldmann}, {Torrey}, {Ma},
  {Angl{\'e}s-Alc{\'a}zar}, {Su}, {Orr}, {Schmitz}, {Escala}, {Sanderson},
  {Grudi{\'c}}, {Hafen}, {Kim}, {Fitts}, {Bullock}, {Wheeler}, {Chan},
  {Elbert}, \& {Narayanan}}]{Hopkins2018MNRAS.480..800H}
{Hopkins}, P.~F., {Wetzel}, A., {Keres}, D., {et~al.} 2018,
  \href{http://dx.doi.org/10.1093/mnras/sty1690}{\color{magenta}\mnras},
  \href{https://ui.adsabs.harvard.edu/abs/2018MNRAS.480..800H}{480, 800}

\bibitem[{Hunter(2007)}]{matplotlib}
Hunter, J.~D. 2007,
  \href{http://dx.doi.org/10.1109/MCSE.2007.55}{\color{magenta}Computing in
  Science Engineering}, 9, 90

\bibitem[{{J{\'a}quez-Dom{\'\i}nguez}
  {et~al.}(2023){J{\'a}quez-Dom{\'\i}nguez}, {Galv{\'a}n-Madrid}, {Fritz},
  {Zamora-Avil{\'e}s}, {Camps}, {Bruzual}, {Baes}, {Lin}, \&
  {V{\'a}zquez-Semadeni}}]{Jaquez-Dominguez2023}
{J{\'a}quez-Dom{\'\i}nguez}, J.~M., {Galv{\'a}n-Madrid}, R., {Fritz}, J.,
  {et~al.} 2023,
  \href{http://dx.doi.org/10.3847/1538-4357/accae7}{\color{magenta}\apj},
  \href{https://ui.adsabs.harvard.edu/abs/2023ApJ...950...88J}{950, 88}

\bibitem[{{Jeffreson} {et~al.}(2024){Jeffreson}, {Semenov}, \&
  {Krumholz}}]{jeffreson:2024}
{Jeffreson}, S. M.~R., {Semenov}, V.~A., \& {Krumholz}, M.~R. 2024,
  \href{http://dx.doi.org/10.1093/mnras/stad3550}{\color{magenta}\mnras},
  \href{https://ui.adsabs.harvard.edu/abs/2024MNRAS.527.7093J}{527, 7093}

\bibitem[{{Kahn}(1954)}]{Kahn1954}
{Kahn}, F.~D. 1954, \bain,
  \href{https://ui.adsabs.harvard.edu/abs/1954BAN....12..187K}{12, 187}

\bibitem[{{Kakiichi} \& {Gronke}(2021)}]{Kakiichi2021ApJ...908...30K}
{Kakiichi}, K. \& {Gronke}, M. 2021,
  \href{http://dx.doi.org/10.3847/1538-4357/abc2d9}{\color{magenta}\apj},
  \href{https://ui.adsabs.harvard.edu/abs/2021ApJ...908...30K}{908, 30}

\bibitem[{{Kannan} {et~al.}(2022){Kannan}, {Garaldi}, {Smith}, {Pakmor},
  {Springel}, {Vogelsberger}, \& {Hernquist}}]{Kannan2022}
{Kannan}, R., {Garaldi}, E., {Smith}, A., {et~al.} 2022,
  \href{http://dx.doi.org/10.1093/mnras/stab3710}{\color{magenta}\mnras},
  \href{https://ui.adsabs.harvard.edu/abs/2022MNRAS.511.4005K}{511, 4005}

\bibitem[{{Kennicutt}(1998)}]{Kennicutt1998ApJ}
{Kennicutt}, Robert~C., J. 1998,
  \href{http://dx.doi.org/10.1086/305588}{\color{magenta}\apj},
  \href{https://ui.adsabs.harvard.edu/abs/1998ApJ...498..541K}{498, 541}

\bibitem[{{Kessel-Deynet} \& {Burkert}(2003)}]{Kessel-Deynet2003}
{Kessel-Deynet}, O. \& {Burkert}, A. 2003,
  \href{http://dx.doi.org/10.1046/j.1365-8711.2003.05737.x}{\color{magenta}\mnras},
  \href{https://ui.adsabs.harvard.edu/abs/2003MNRAS.338..545K}{338, 545}

\bibitem[{{Kim} \& {Ostriker}(2017)}]{kim:2017}
{Kim}, C.-G. \& {Ostriker}, E.~C. 2017,
  \href{http://dx.doi.org/10.3847/1538-4357/aa8599}{\color{magenta}\apj},
  \href{https://ui.adsabs.harvard.edu/abs/2017ApJ...846..133K}{846, 133}

\bibitem[{{Koepferl} \& {Robitaille}(2017)}]{Koepferl2017ApJ...849....3K}
{Koepferl}, C.~M. \& {Robitaille}, T.~P. 2017,
  \href{http://dx.doi.org/10.3847/1538-4357/aa8666}{\color{magenta}\apj},
  \href{https://ui.adsabs.harvard.edu/abs/2017ApJ...849....3K}{849, 3}

\bibitem[{{Kroupa}(2001)}]{Kroupa2001MNRAS}
{Kroupa}, P. 2001,
  \href{http://dx.doi.org/10.1046/j.1365-8711.2001.04022.x}{\color{magenta}\mnras},
  \href{https://ui.adsabs.harvard.edu/abs/2001MNRAS.322..231K}{322, 231}

\bibitem[{{Kroupa}(2002)}]{Kroupa2002}
{Kroupa}, P. 2002,
  \href{http://dx.doi.org/10.1126/science.1067524}{\color{magenta}Science},
  \href{https://ui.adsabs.harvard.edu/abs/2002Sci...295...82K}{295, 82}

\bibitem[{{Krumholz} {et~al.}(2012){Krumholz}, {Dekel}, \&
  {McKee}}]{Krumholz2012ApJ...745...69K}
{Krumholz}, M.~R., {Dekel}, A., \& {McKee}, C.~F. 2012,
  \href{http://dx.doi.org/10.1088/0004-637X/745/1/69}{\color{magenta}\apj},
  \href{https://ui.adsabs.harvard.edu/abs/2012ApJ...745...69K}{745, 69}

\bibitem[{{Ladjelate} {et~al.}(2020){Ladjelate}, {Andr{\'e}}, {K{\"o}nyves},
  {Ward-Thompson}, {Men'shchikov}, {Bracco}, {Palmeirim}, {Roy}, {Shimajiri},
  {Kirk}, {Arzoumanian}, {Benedettini}, {Di Francesco}, {Fiorellino},
  {Schneider}, {Pezzuto}, {Motte}, \& {Herschel Gould Belt Survey
  Team}}]{Ladjelate2020}
{Ladjelate}, B., {Andr{\'e}}, P., {K{\"o}nyves}, V., {et~al.} 2020,
  \href{http://dx.doi.org/10.1051/0004-6361/201936442}{\color{magenta}\aap},
  \href{https://ui.adsabs.harvard.edu/abs/2020A&A...638A..74L}{638, A74}

\bibitem[{{Lam} {et~al.}(2015){Lam}, {Pitrou}, \& {Seibert}}]{numba}
{Lam}, S.~K., {Pitrou}, A., \& {Seibert}, S. 2015, in Proc. Second Workshop on
  the LLVM Compiler Infrastructure in HPC,
  \href{https://ui.adsabs.harvard.edu/abs/2015llvm.confE...1L}{1--6}

\bibitem[{{Lancaster} {et~al.}(2021{\natexlab{a}}){Lancaster}, {Ostriker},
  {Kim}, \& {Kim}}]{lancaster:2021}
{Lancaster}, L., {Ostriker}, E.~C., {Kim}, J.-G., \& {Kim}, C.-G.
  2021{\natexlab{a}},
  \href{http://dx.doi.org/10.3847/1538-4357/abf8ac}{\color{magenta}\apj},
  \href{https://ui.adsabs.harvard.edu/abs/2021ApJ...914...90L}{914, 90}

\bibitem[{{Lancaster} {et~al.}(2021{\natexlab{b}}){Lancaster}, {Ostriker},
  {Kim}, \& {Kim}}]{lancaster:2021_b}
{Lancaster}, L., {Ostriker}, E.~C., {Kim}, J.-G., \& {Kim}, C.-G.
  2021{\natexlab{b}},
  \href{http://dx.doi.org/10.3847/2041-8213/ac3333}{\color{magenta}\apjl},
  \href{https://ui.adsabs.harvard.edu/abs/2021ApJ...922L...3L}{922, L3}

\bibitem[{{Lee} {et~al.}(2016){Lee}, {Miville-Desch{\^e}nes}, \&
  {Murray}}]{Lee2016ApJ...833..229L}
{Lee}, E.~J., {Miville-Desch{\^e}nes}, M.-A., \& {Murray}, N.~W. 2016,
  \href{http://dx.doi.org/10.3847/1538-4357/833/2/229}{\color{magenta}\apj},
  \href{https://ui.adsabs.harvard.edu/abs/2016ApJ...833..229L}{833, 229}

\bibitem[{{Leitherer} {et~al.}(1999){Leitherer}, {Schaerer}, {Goldader},
  {Delgado}, {Robert}, {Kune}, {de Mello}, {Devost}, \&
  {Heckman}}]{Leitherer1999ApJS}
{Leitherer}, C., {Schaerer}, D., {Goldader}, J.~D., {et~al.} 1999,
  \href{http://dx.doi.org/10.1086/313233}{\color{magenta}\apjs},
  \href{https://ui.adsabs.harvard.edu/abs/1999ApJS..123....3L}{123, 3}

\bibitem[{{Liang} {et~al.}(2021){Liang}, {Feldmann}, {Hayward}, {Narayanan},
  {Catmabacak}, {Keres}, {Faucher-Gigu{\`e}re}, \& {Hopkins}}]{Liang:IRXbeta}
{Liang}, L., {Feldmann}, R., {Hayward}, C.~C., {et~al.} 2021,
  \href{http://dx.doi.org/10.1093/mnras/stab096}{\color{magenta}\mnras},
  \href{https://ui.adsabs.harvard.edu/abs/2021MNRAS.502.3210L}{502, 3210}

\bibitem[{{Lin} {et~al.}(2017){Lin}, {Liu}, {Dale}, {Li}, {Busquet}, {Zhang},
  {Ginsburg}, {Galv{\'a}n-Madrid}, {Kov{\'a}cs}, {Koch}, {Qian}, {Wang},
  {Longmore}, {Chen}, \& {Walker}}]{Lin2017ApJ...840...22L}
{Lin}, Y., {Liu}, H.~B., {Dale}, J.~E., {et~al.} 2017,
  \href{http://dx.doi.org/10.3847/1538-4357/aa6c67}{\color{magenta}\apj},
  \href{https://ui.adsabs.harvard.edu/abs/2017ApJ...840...22L}{840, 22}

\bibitem[{{Louvet} {et~al.}(2014){Louvet}, {Motte}, {Hennebelle}, {Maury},
  {Bonnell}, {Bontemps}, {Gusdorf}, {Hill}, {Gueth}, {Peretto},
  {Duarte-Cabral}, {Stephan}, {Schilke}, {Csengeri}, {Nguyen Luong}, \&
  {Lis}}]{Louvet2014}
{Louvet}, F., {Motte}, F., {Hennebelle}, P., {et~al.} 2014,
  \href{http://dx.doi.org/10.1051/0004-6361/201423603}{\color{magenta}\aap},
  \href{https://ui.adsabs.harvard.edu/abs/2014A&A...570A..15L}{570, A15}

\bibitem[{{Lovell} {et~al.}(2021){Lovell}, {Vijayan}, {Thomas}, {Wilkins},
  {Barnes}, {Irodotou}, \& {Roper}}]{Lovell2021MNRAS}
{Lovell}, C.~C., {Vijayan}, A.~P., {Thomas}, P.~A., {et~al.} 2021,
  \href{http://dx.doi.org/10.1093/mnras/staa3360}{\color{magenta}\mnras},
  \href{https://ui.adsabs.harvard.edu/abs/2021MNRAS.500.2127L}{500, 2127}

\bibitem[{{Matzner}(2002)}]{Matzner2002ApJ}
{Matzner}, C.~D. 2002,
  \href{http://dx.doi.org/10.1086/338030}{\color{magenta}\apj},
  \href{https://ui.adsabs.harvard.edu/abs/2002ApJ...566..302M}{566, 302}

\bibitem[{{Menon} {et~al.}(2023){Menon}, {Federrath}, \&
  {Krumholz}}]{menon:2023}
{Menon}, S.~H., {Federrath}, C., \& {Krumholz}, M.~R. 2023,
  \href{http://dx.doi.org/10.1093/mnras/stad856}{\color{magenta}\mnras},
  \href{https://ui.adsabs.harvard.edu/abs/2023MNRAS.521.5160M}{521, 5160}

\bibitem[{{Meurer} {et~al.}(1999){Meurer}, {Heckman}, \&
  {Calzetti}}]{Meurer1999ApJ}
{Meurer}, G.~R., {Heckman}, T.~M., \& {Calzetti}, D. 1999,
  \href{http://dx.doi.org/10.1086/307523}{\color{magenta}\apj},
  \href{https://ui.adsabs.harvard.edu/abs/1999ApJ...521...64M}{521, 64}

\bibitem[{{Molinari} {et~al.}(2010){Molinari}, {Swinyard}, {Bally}, {Barlow},
  {Bernard}, {Martin}, {Moore}, {Noriega-Crespo}, {Plume}, {Testi}, {Zavagno},
  {Abergel}, {Ali}, {Anderson}, {Andr{\'e}}, {Baluteau}, {Battersby},
  {Beltr{\'a}n}, {Benedettini}, {Billot}, {Blommaert}, {Bontemps}, {Boulanger},
  {Brand}, {Brunt}, {Burton}, {Calzoletti}, {Carey}, {Caselli}, {Cesaroni},
  {Cernicharo}, {Chakrabarti}, {Chrysostomou}, {Cohen}, {Compiegne}, {de
  Bernardis}, {de Gasperis}, {di Giorgio}, {Elia}, {Faustini}, {Flagey},
  {Fukui}, {Fuller}, {Ganga}, {Garcia-Lario}, {Glenn}, {Goldsmith}, {Griffin},
  {Hoare}, {Huang}, {Ikhenaode}, {Joblin}, {Joncas}, {Juvela}, {Kirk},
  {Lagache}, {Li}, {Lim}, {Lord}, {Marengo}, {Marshall}, {Masi}, {Massi},
  {Matsuura}, {Minier}, {Miville-Desch{\^e}nes}, {Montier}, {Morgan}, {Motte},
  {Mottram}, {M{\"u}ller}, {Natoli}, {Neves}, {Olmi}, {Paladini}, {Paradis},
  {Parsons}, {Peretto}, {Pestalozzi}, {Pezzuto}, {Piacentini}, {Piazzo},
  {Polychroni}, {Pomar{\`e}s}, {Popescu}, {Reach}, {Ristorcelli}, {Robitaille},
  {Robitaille}, {Rod{\'o}n}, {Roy}, {Royer}, {Russeil}, {Saraceno}, {Sauvage},
  {Schilke}, {Schisano}, {Schneider}, {Schuller}, {Schulz}, {Sibthorpe},
  {Smith}, {Smith}, {Spinoglio}, {Stamatellos}, {Strafella}, {Stringfellow},
  {Sturm}, {Taylor}, {Thompson}, {Traficante}, {Tuffs}, {Umana}, {Valenziano},
  {Vavrek}, {Veneziani}, {Viti}, {Waelkens}, {Ward-Thompson}, {White},
  {Wilcock}, {Wyrowski}, {Yorke}, \& {Zhang}}]{Molinari2010}
{Molinari}, S., {Swinyard}, B., {Bally}, J., {et~al.} 2010,
  \href{http://dx.doi.org/10.1051/0004-6361/201014659}{\color{magenta}\aap},
  \href{https://ui.adsabs.harvard.edu/abs/2010A&A...518L.100M}{518, L100}

\bibitem[{Ostriker \& McKee(1988)}]{Ostriker1988}
Ostriker, J.~P. \& McKee, C.~F. 1988,
  \href{http://dx.doi.org/10.1103/RevModPhys.60.1}{\color{magenta}Rev. Mod.
  Phys.}, 60, 1

\bibitem[{{Overzier} {et~al.}(2011){Overzier}, {Heckman}, {Wang}, {Armus},
  {Buat}, {Howell}, {Meurer}, {Seibert}, {Siana}, {Basu-Zych}, {Charlot},
  {Gon{\c{c}}alves}, {Martin}, {Neill}, {Rich}, {Salim}, \&
  {Schiminovich}}]{Overzier2011ApJ}
{Overzier}, R.~A., {Heckman}, T.~M., {Wang}, J., {et~al.} 2011,
  \href{http://dx.doi.org/10.1088/2041-8205/726/1/L7}{\color{magenta}\apjl},
  \href{https://ui.adsabs.harvard.edu/abs/2011ApJ...726L...7O}{726, L7}

\bibitem[{{Pallottini} \& {Ferrara}(2023)}]{pallottini:2023}
{Pallottini}, A. \& {Ferrara}, A. 2023,
  \href{http://dx.doi.org/10.1051/0004-6361/202347384}{\color{magenta}\aap},
  \href{https://ui.adsabs.harvard.edu/abs/2023A&A...677L...4P}{677, L4}

\bibitem[{{Pallottini} {et~al.}(2017){Pallottini}, {Ferrara}, {Bovino},
  {Vallini}, {Gallerani}, {Maiolino}, \& {Salvadori}}]{pallottini:2017}
{Pallottini}, A., {Ferrara}, A., {Bovino}, S., {et~al.} 2017,
  \href{http://dx.doi.org/10.1093/mnras/stx1792}{\color{magenta}\mnras},
  \href{https://ui.adsabs.harvard.edu/abs/2017MNRAS.471.4128P}{471, 4128}

\bibitem[{{Pallottini} {et~al.}(2019){Pallottini}, {Ferrara}, {Decataldo},
  {Gallerani}, {Vallini}, {Carniani}, {Behrens}, {Kohandel}, \&
  {Salvadori}}]{pallottini:2019}
{Pallottini}, A., {Ferrara}, A., {Decataldo}, D., {et~al.} 2019,
  \href{http://dx.doi.org/10.1093/mnras/stz1383}{\color{magenta}\mnras},
  \href{https://ui.adsabs.harvard.edu/abs/2019MNRAS.487.1689P}{487, 1689}

\bibitem[{{Pallottini} {et~al.}(2022){Pallottini}, {Ferrara}, {Gallerani},
  {Behrens}, {Kohandel}, {Carniani}, {Vallini}, {Salvadori}, {Gelli},
  {Sommovigo}, {D'Odorico}, {Di Mascia}, \& {Pizzati}}]{pallottini:2022}
{Pallottini}, A., {Ferrara}, A., {Gallerani}, S., {et~al.} 2022,
  \href{http://dx.doi.org/10.1093/mnras/stac1281}{\color{magenta}\mnras},
  \href{https://ui.adsabs.harvard.edu/abs/2022MNRAS.513.5621P}{513, 5621}

\bibitem[{{Pillepich} {et~al.}(2018){Pillepich}, {Springel}, {Nelson}, {Genel},
  {Naiman}, {Pakmor}, {Hernquist}, {Torrey}, {Vogelsberger}, {Weinberger}, \&
  {Marinacci}}]{Pillepich2018}
{Pillepich}, A., {Springel}, V., {Nelson}, D., {et~al.} 2018,
  \href{http://dx.doi.org/10.1093/mnras/stx2656}{\color{magenta}\mnras},
  \href{https://ui.adsabs.harvard.edu/abs/2018MNRAS.473.4077P}{473, 4077}

\bibitem[{{Pontzen} {et~al.}(2013){Pontzen}, {Rovskar}, {Stinson}, {Woods},
  {Reed}, {Coles}, \& {Quinn}}]{pynbody}
{Pontzen}, A., {Rovskar}, R., {Stinson}, G.~S., {et~al.} 2013, {pynbody:
  Astrophysics Simulation Analysis for Python}, astrophysics Source Code
  Library, ascl:1305.002

\bibitem[{{Potdar} {et~al.}(2022){Potdar}, {Das}, {Issac}, {Tej}, {Vig}, \&
  {Chandra}}]{Potdar2022MNRAS.510..658P}
{Potdar}, A., {Das}, S.~R., {Issac}, N., {et~al.} 2022,
  \href{http://dx.doi.org/10.1093/mnras/stab3479}{\color{magenta}\mnras},
  \href{https://ui.adsabs.harvard.edu/abs/2022MNRAS.510..658P}{510, 658}

\bibitem[{{Reddy} {et~al.}(2015){Reddy}, {Kriek}, {Shapley}, {Freeman},
  {Siana}, {Coil}, {Mobasher}, {Price}, {Sanders}, \& {Shivaei}}]{Reddy2015ApJ}
{Reddy}, N.~A., {Kriek}, M., {Shapley}, A.~E., {et~al.} 2015,
  \href{http://dx.doi.org/10.1088/0004-637X/806/2/259}{\color{magenta}\apj},
  \href{https://ui.adsabs.harvard.edu/abs/2015ApJ...806..259R}{806, 259}

\bibitem[{{Reddy} {et~al.}(2018){Reddy}, {Oesch}, {Bouwens}, {Montes},
  {Illingworth}, {Steidel}, {van Dokkum}, {Atek}, {Carollo}, {Cibinel},
  {Holden}, {Labb{\'e}}, {Magee}, {Morselli}, {Nelson}, \&
  {Wilkins}}]{Reddy2018ApJ}
{Reddy}, N.~A., {Oesch}, P.~A., {Bouwens}, R.~J., {et~al.} 2018,
  \href{http://dx.doi.org/10.3847/1538-4357/aaa3e7}{\color{magenta}\apj},
  \href{https://ui.adsabs.harvard.edu/abs/2018ApJ...853...56R}{853, 56}

\bibitem[{{Reissl} {et~al.}(2020){Reissl}, {Stil}, {Chen}, {Tre{\ss}},
  {Sormani}, {Smith}, {Klessen}, {Buick}, {Glover}, {Shanahan}, {Lemmer},
  {Soler}, {Beuther}, {Urquhart}, {Anderson}, {Menten}, {Brunthaler}, {Ragan},
  \& {Rugel}}]{Reissl2020A&A...642A.201R}
{Reissl}, S., {Stil}, J.~M., {Chen}, E., {et~al.} 2020,
  \href{http://dx.doi.org/10.1051/0004-6361/202037690}{\color{magenta}\aap},
  \href{https://ui.adsabs.harvard.edu/abs/2020A&A...642A.201R}{642, A201}

\bibitem[{{Rosdahl} {et~al.}(2013){Rosdahl}, {Blaizot}, {Aubert}, {Stranex}, \&
  {Teyssier}}]{Rosdahl2013MNRAS}
{Rosdahl}, J., {Blaizot}, J., {Aubert}, D., {Stranex}, T., \& {Teyssier}, R.
  2013, \href{http://dx.doi.org/10.1093/mnras/stt1722}{\color{magenta}\mnras},
  \href{https://ui.adsabs.harvard.edu/abs/2013MNRAS.436.2188R}{436, 2188}

\bibitem[{{Schaller} {et~al.}(2015){Schaller}, {Dalla Vecchia}, {Schaye},
  {Bower}, {Theuns}, {Crain}, {Furlong}, \& {McCarthy}}]{Schaller2015}
{Schaller}, M., {Dalla Vecchia}, C., {Schaye}, J., {et~al.} 2015,
  \href{http://dx.doi.org/10.1093/mnras/stv2169}{\color{magenta}\mnras},
  \href{https://ui.adsabs.harvard.edu/abs/2015MNRAS.454.2277S}{454, 2277}

\bibitem[{{Schmidt}(1959)}]{Schmidt1959ApJ}
{Schmidt}, M. 1959,
  \href{http://dx.doi.org/10.1086/146614}{\color{magenta}\apj},
  \href{https://ui.adsabs.harvard.edu/abs/1959ApJ...129..243S}{129, 243}

\bibitem[{Sedov(1958)}]{Sedov1958}
Sedov, L.~I. 1958,
  \href{http://dx.doi.org/10.1103/RevModPhys.30.1077}{\color{magenta}Rev. Mod.
  Phys.}, 30, 1077

\bibitem[{{Smith} {et~al.}(2020){Smith}, {Tre{\ss}}, {Sormani}, {Glover},
  {Klessen}, {Clark}, {Izquierdo}, {Duarte-Cabral}, \&
  {Zucker}}]{Smith2020MNRAS.492.1594S}
{Smith}, R.~J., {Tre{\ss}}, R.~G., {Sormani}, M.~C., {et~al.} 2020,
  \href{http://dx.doi.org/10.1093/mnras/stz3328}{\color{magenta}\mnras},
  \href{https://ui.adsabs.harvard.edu/abs/2020MNRAS.492.1594S}{492, 1594}

\bibitem[{{Sommovigo} {et~al.}(2020){Sommovigo}, {Ferrara}, {Pallottini},
  {Carniani}, {Gallerani}, \& {Decataldo}}]{Sommovigo:2020}
{Sommovigo}, L., {Ferrara}, A., {Pallottini}, A., {et~al.} 2020,
  \href{http://dx.doi.org/10.1093/mnras/staa1959}{\color{magenta}\mnras},
  \href{https://ui.adsabs.harvard.edu/abs/2020MNRAS.497..956S}{497, 956}

\bibitem[{{Sun} {et~al.}(2023){Sun}, {Faucher-Gigu{\`e}re}, {Hayward}, {Shen},
  {Wetzel}, \& {Cochrane}}]{Sun2023ApJ...955L..35S}
{Sun}, G., {Faucher-Gigu{\`e}re}, C.-A., {Hayward}, C.~C., {et~al.} 2023,
  \href{http://dx.doi.org/10.3847/2041-8213/acf85a}{\color{magenta}\apjl},
  \href{https://ui.adsabs.harvard.edu/abs/2023ApJ...955L..35S}{955, L35}

\bibitem[{{Takeuchi} {et~al.}(2012){Takeuchi}, {Yuan}, {Ikeyama}, {Murata}, \&
  {Inoue}}]{Takeuchi2012ApJ}
{Takeuchi}, T.~T., {Yuan}, F.-T., {Ikeyama}, A., {Murata}, K.~L., \& {Inoue},
  A.~K. 2012,
  \href{http://dx.doi.org/10.1088/0004-637X/755/2/144}{\color{magenta}\apj},
  \href{https://ui.adsabs.harvard.edu/abs/2012ApJ...755..144T}{755, 144}

\bibitem[{{Teyssier}(2002)}]{Teyssier2002}
{Teyssier}, R. 2002,
  \href{http://dx.doi.org/10.1051/0004-6361:20011817}{\color{magenta}\aap},
  \href{https://ui.adsabs.harvard.edu/abs/2002A&A...385..337T}{385, 337}

\bibitem[{{Trayford} {et~al.}(2017){Trayford}, {Camps}, {Theuns}, {Baes},
  {Bower}, {Crain}, {Gunawardhana}, {Schaller}, {Schaye}, \&
  {Frenk}}]{Trayford2017MNRAS}
{Trayford}, J.~W., {Camps}, P., {Theuns}, T., {et~al.} 2017,
  \href{http://dx.doi.org/10.1093/mnras/stx1051}{\color{magenta}\mnras},
  \href{https://ui.adsabs.harvard.edu/abs/2017MNRAS.470..771T}{470, 771}

\bibitem[{van~der Walt {et~al.}(2011)van~der Walt, Colbert, \&
  Varoquaux}]{numpy}
van~der Walt, S., Colbert, S.~C., \& Varoquaux, G. 2011,
  \href{http://dx.doi.org/10.1109/MCSE.2011.37}{\color{magenta}Computing in
  Science Engineering}, 13, 22

\bibitem[{Van~Rossum \& de~Boer(1991)}]{python2}
Van~Rossum, G. \& de~Boer, J. 1991, CWI Quarterly, 4, 283

\bibitem[{Van~Rossum \& Drake(2009)}]{python3}
Van~Rossum, G. \& Drake, F.~L. 2009, Python 3 Reference Manual (Scotts Valley,
  CA: CreateSpace)

\bibitem[{{V{\'a}zquez-Semadeni} {et~al.}(2019){V{\'a}zquez-Semadeni}, {Palau},
  {Ballesteros-Paredes}, {G{\'o}mez}, \&
  {Zamora-Avil{\'e}s}}]{Vazquez-Semadeni2019}
{V{\'a}zquez-Semadeni}, E., {Palau}, A., {Ballesteros-Paredes}, J.,
  {G{\'o}mez}, G.~C., \& {Zamora-Avil{\'e}s}, M. 2019,
  \href{http://dx.doi.org/10.1093/mnras/stz2736}{\color{magenta}\mnras},
  \href{https://ui.adsabs.harvard.edu/abs/2019MNRAS.490.3061V}{490, 3061}

\bibitem[{{Viaene} {et~al.}(2020){Viaene}, {Nersesian}, {Fritz}, {Verstocken},
  {Baes}, {Bianchi}, {Casasola}, {Cassar{\`a}}, {Clark}, {Davies}, {De Looze},
  {De Vis}, {Dobbels}, {Galametz}, {Galliano}, {Jones}, {Madden}, {Mosenkov},
  {Trcka}, {Xilouris}, \& {Ysard}}]{viaene:2020}
{Viaene}, S., {Nersesian}, A., {Fritz}, J., {et~al.} 2020,
  \href{http://dx.doi.org/10.1051/0004-6361/202037476}{\color{magenta}\aap},
  \href{https://ui.adsabs.harvard.edu/abs/2020A&A...638A.150V}{638, A150}

\bibitem[{{Vijayan} {et~al.}(2022){Vijayan}, {Wilkins}, {Lovell}, {Thomas},
  {Camps}, {Baes}, {Trayford}, {Kuusisto}, \& {Roper}}]{Vijayan2022FLARESIII}
{Vijayan}, A.~P., {Wilkins}, S.~M., {Lovell}, C.~C., {et~al.} 2022,
  \href{http://dx.doi.org/10.1093/mnras/stac338}{\color{magenta}\mnras},
  \href{https://ui.adsabs.harvard.edu/abs/2022MNRAS.511.4999V}{511, 4999}

\bibitem[{{Virtanen} {et~al.}(2020){Virtanen}, {Gommers}, {Oliphant},
  {Haberland}, {Reddy}, {Cournapeau}, {Burovski}, {Peterson}, {Weckesser},
  {Bright}, {van der Walt}, {Brett}, {Wilson}, {Millman}, {Mayorov}, {Nelson},
  {Jones}, {Kern}, {Larson}, {Carey}, {Polat}, {Feng}, {Moore}, {VanderPlas},
  {Laxalde}, {Perktold}, {Cimrman}, {Henriksen}, {Quintero}, {Harris},
  {Archibald}, {Ribeiro}, {Pedregosa}, {van Mulbregt}, \& {SciPy 1. 0
  Contributors}}]{scipy}
{Virtanen}, P., {Gommers}, R., {Oliphant}, T.~E., {et~al.} 2020,
  \href{http://dx.doi.org/10.1038/s41592-019-0686-2}{\color{magenta}Nature
  Methods}, \href{https://ui.adsabs.harvard.edu/abs/2020NatMe..17..261V}{17,
  261}

\bibitem[{{Weaver} {et~al.}(1977){Weaver}, {McCray}, {Castor}, {Shapiro}, \&
  {Moore}}]{Weaver1977ApJ}
{Weaver}, R., {McCray}, R., {Castor}, J., {Shapiro}, P., \& {Moore}, R. 1977,
  \href{http://dx.doi.org/10.1086/155692}{\color{magenta}\apj},
  \href{https://ui.adsabs.harvard.edu/abs/1977ApJ...218..377W}{218, 377}

\bibitem[{{Weingartner} \& {Draine}(2001)}]{Weingartner:2001}
{Weingartner}, J.~C. \& {Draine}, B.~T. 2001,
  \href{http://dx.doi.org/10.1086/318651}{\color{magenta}\apj},
  \href{https://ui.adsabs.harvard.edu/abs/2001ApJ...548..296W}{548, 296}

\bibitem[{{Whitworth}(1979)}]{Whitworth1979}
{Whitworth}, A. 1979,
  \href{http://dx.doi.org/10.1093/mnras/186.1.59}{\color{magenta}\mnras},
  \href{https://ui.adsabs.harvard.edu/abs/1979MNRAS.186...59W}{186, 59}

\bibitem[{{Williams} \& {McKee}(1997)}]{Williams1997ApJ...476..166W}
{Williams}, J.~P. \& {McKee}, C.~F. 1997,
  \href{http://dx.doi.org/10.1086/303588}{\color{magenta}\apj},
  \href{https://ui.adsabs.harvard.edu/abs/1997ApJ...476..166W}{476, 166}

\bibitem[{{Williams} {et~al.}(2018){Williams}, {Bisbas}, {Haworth}, \&
  {Mackey}}]{Williams2018}
{Williams}, R. J.~R., {Bisbas}, T.~G., {Haworth}, T.~J., \& {Mackey}, J. 2018,
  \href{http://dx.doi.org/10.1093/mnras/sty1484}{\color{magenta}\mnras},
  \href{https://ui.adsabs.harvard.edu/abs/2018MNRAS.479.2016W}{479, 2016}

\bibitem[{{Zamora-Avil{\'e}s} {et~al.}(2019){Zamora-Avil{\'e}s},
  {V{\'a}zquez-Semadeni}, {Gonz{\'a}lez}, {Franco}, {Shore}, {Hartmann},
  {Ballesteros-Paredes}, {Banerjee}, \&
  {K{\"o}rtgen}}]{Zamora2019MNRAS.487.2200Z}
{Zamora-Avil{\'e}s}, M., {V{\'a}zquez-Semadeni}, E., {Gonz{\'a}lez}, R.~F.,
  {et~al.} 2019,
  \href{http://dx.doi.org/10.1093/mnras/stz1235}{\color{magenta}\mnras},
  \href{https://ui.adsabs.harvard.edu/abs/2019MNRAS.487.2200Z}{487, 2200}

\end{thebibliography}
